\documentclass[prl,
reprint,
twocolumn,
aps,
superscriptaddress,
article]{revtex4-2}
\usepackage{epsfig}
\usepackage{graphicx}
\usepackage{dcolumn}
\usepackage{hyperref}
\usepackage{ulem}
\usepackage{bm}
\usepackage{color}
\usepackage{braket}
\usepackage{latexsym}
\usepackage{todonotes}
\usepackage{amsmath}
\usepackage{amssymb}
\hypersetup{bookmarks,
            colorlinks,
           citecolor=red,
            filecolor=blue,
           urlcolor=blue,
           pdfpagemode=None}
\usepackage{latexsym}
\usepackage{physics}
\usepackage{lipsum}
\usepackage{lineno}
\usepackage{multirow}
\def \beq{\begin{eqnarray}}
\def \eeq{\end{eqnarray}}

\def \r{{\bm g}}

\def \m{{\mathbf m}}
\def \a{{\bm a}}
\def \b{{\bm b}}
\def \r{{\bm r}}

\def \k{{\bm k}}

\def \q{{\bm q}}

\def \M{{\bm M}}

\newcounter{para}

\begin{document}
\title{Fluctuation-driven chiral ferromagnetism}
\author{Rokas Veitas}
\thanks{R.V. and A.K. contributed equally to this work.}
\affiliation{Department of Physics, Carnegie Mellon University, Pittsburgh, PA 15213, USA}
\author{Ahmed Khalifa}
\thanks{R.V. and A.K. contributed equally to this work.}
\affiliation{Department of Physics, Carnegie Mellon University, Pittsburgh, PA 15213, USA}
\author{Francisco Machado}
\affiliation{ QuTech, Delft University of Technology, PO Box 5046, 2600 GA Delft, The Netherlands}
\author{Shubhayu Chatterjee}
\affiliation{Department of Physics, Carnegie Mellon University, Pittsburgh, PA 15213, USA}

\begin{abstract}
Quantum fluctuations are often suppressed in ferromagnetic materials because they admit a simple unfrustrated ground state, greatly limiting the scope of the phenomena that can be observed.
In this work, we show how naturally occurring magnetization-non-conserving couplings fundamentally alter this paradigm by demonstrating the existence of a chiral ferromagnet that is stabilized by quantum fluctuations.
More specifically, we show how these spin-orbit interactions modify the classical phase diagram; whereas a classical analysis predicts only achiral collinear states, we observe fluctuation-stabilized phases, including a ferromagnet with large orbital chirality and a chiral stripe.
We elucidate how such couplings generate a scalar orbital chirality spontaneously, in contrast to conventional mechanisms which rely upon a field-induced canting of vector chiral order.
The resultant chiral states exhibit distinct transport signatures, namely an enhanced thermal Hall effect, and are of direct relevance to moiré heterostructures, Rydberg-atom arrays, and solid-state materials featuring non-Kramers spins.
\end{abstract}

\maketitle

Quantum fluctuations are instrumental to generate and stabilize complex correlated electronic phenomena, whether in the context of charge, spin, or orbital degrees of freedom~\cite{sachdev2023quantum,fradkin2021quantum,sachdev:2011}.
However, such fluctuations are strongly suppressed when a system admits a simple, classical ground state configuration~\cite{Auerbach_Book}. 
Consequently, diagnosing regimes where large quantum fluctuations can emerge within classically ordered phases constitutes a key challenge in quantum many-body physics as it serves as a precursor to uncovering qualitatively novel phenomena.

One paradigmatic setting of particular interest are ferromagnets.
Owing to a lack of frustration, they admit simple classical ground states, which has limited exploration of quantum fluctuation driven phenomena. 
This contrasts sharply with antiferromagnets (AFM), where even nearest-neighbor interactions on bipartite lattices induce quantum fluctuations, further enhanced when the system exhibits geometric frustration or longer-ranged interactions~\cite{ramirez1994strongly,lacroix2011introduction,Sondhi2008}. 
Such quantum fluctuations enable fascinating correlated phenomena in AFMs, ranging from entangled spin liquids~\cite{Savary2016QSL,Zhou2017QSL,Broholm2020QSL,Clark2021QSL} to deconfined quantum criticality~\cite{senthil2004deconfined,ChongWangDQCP2017,senthil2024deconfined}. 
In ferromagnets (FM), modifying lattice geometry or interaction range leaves the classical ground state intact, rendering such routes ineffective.
Establishing mechanisms that can induce and control strong quantum fluctuations in FM systems therefore  holds promise to design and realize unexplored quantum phases.

\begin{figure}
\centering
\includegraphics{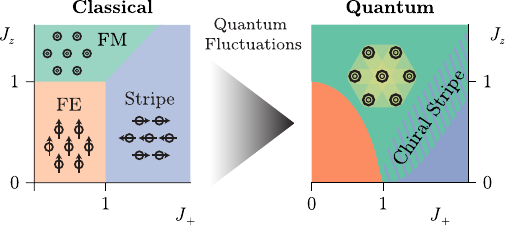}
    \caption{
The mean-field classical phase diagram is strongly modified by quantum fluctuations: the FM phase dominates along the classical FE-stripe boundary, even at very small $J_z$. 
In this ferromagnetic region, fluctuations induce a non-zero orbital chirality.
Between the FM and stripe phases, we see evidence of a distinct chiral stripe phase.}
    \label{fig:phasediagram}
\end{figure}

Motivated by current advances in quantum simulators across both condensed matter and ultracold atomic platforms, we demonstrate how spin-orbit interactions can induce strong quantum fluctuations in conventional ferromagnets, leading to novel non-perturbative effects. 
Specifically, symmetry-constrained spin-orbit coupling terms promote the spontaneous generation of scalar orbital chirality, even when the underlying classical ordering patterns are strictly collinear~[Fig.~\ref{fig:phasediagram}]. 
Remarkably, the resulting scalar chirality does not originate from the conventional mechanism of combining vector chirality and field-induced magnetization, but rather from quantum fluctuations generated by spin non-conserving terms. 
The resultant state exhibits a non-zero thermal Hall conductance. 
Finally, we argue that such quantum fluctuation stabilized phases are broadly accessible by outlining concrete implementations in moir\'e heterostructures, neutral-atom arrays, and solid-state materials hosting non-Kramers spins. 

\textit{Model and Symmetries.---}
To illustrate our proposal in the simplest context, let us consider a nearest-neighbor (NN) ferromagnetic XXZ Hamiltonian on a triangular lattice, where the pseudospin-up and -down states are encoded in orbital angular momentum degrees of freedom (we denote the local pseudospin operator with $\bm\sigma$).  
The proposed orbital encoding has two important effects \cite{KVMC2026,Liu2018}. 
First, while $\sigma^z$ is flipped under time-reversal (i.e. is odd), the in-plane projection of the pseudospin $(\sigma^x, \sigma^y)$ transforms trivially (i.e. is even) under the same symmetry action.
As a result, out-of-plane ordering corresponds to ferromagnetism, while in-plane ordering corresponds to ferroelectricity, enabling novel multiferroic phases.
Second, such orbital degrees of freedom often arise from on-site potentials with a smaller threefold symmetry~\cite{KVMC2026}, naturally reducing symmetry of in-plane $\bm{\sigma}$-rotations to $C_{3v}$.
The resulting NN Hamiltonian takes the form 
\begin{equation}
\begin{split}
H &= - \sum_{\langle ij \rangle} \bigg[ J_{z} \sigma^z_i \sigma^z_j + \left(\frac{J}{2} \sigma^+_i \sigma^-_j +\mbox{h.c.} \right) \bigg] \\
  &\quad - \sum_{\langle ij \rangle}  J_+( \gamma_{ij}  \, \sigma^+_i \sigma^+_j + \mbox{h.c.} ),
\label{eq:H}
\end{split}
\end{equation}
where the first two terms correspond to a simple ferromagnetic Hamiltonian with easy-axis and easy-plane anisotropy, respectively. 
Henceforth, we set $J=1$.

In this work, we focus on the third term $J_{+}$ which naturally dresses the easy-axis ferromagnet with quantum fluctuations.  
Notably, it is the only bond-dependent coupling allowed by the $C_{3v}$ point group; this symmetry also fixes the phase $\gamma_{ij} = e^{-i2\phi_{ij}}$, where $\phi_{ij}$ is the angle made by the bond $\langle ij \rangle$ with the $+\hat{x}$-coordinate axis. 
This sign structure implies that flipping the sign of $J_+$ is equivalent to a pseudospin rotation of $\pi/2$ about $\hat{z}$; for this reason we restrict our analysis to $J_+ \geq 0$.

By contrast, interaction terms that mix between in- and out-of-plane components (e.g.~$\sigma^+ \sigma^z$), often studied in the context of spin Hamiltonians with spin-orbit coupling~\cite{PhysRevLett.115.167203,PhysRevB.94.174424,Liu2018,Iaconis2018spin,Maksimov2019}, are strictly prohibited by time-reversal.
We note that, while the lack of inversion symmetry of $C_{3v}$ permits a Dzyaloshinskii-Moriya (DM) interaction, we have found that a weak DM term does not significantly affect the quantum phase diagram so we focus on the regime where $\Im[J] =0$~\cite{SM}. 
This condition is enforced if the system preserves $C_6$ symmetry.

\textit{Collinear classical ordering.---}
We begin by considering the classical zero-temperature phase diagram of $H$. 
When $J_+$ is absent, the Hamiltonian $H$ has no frustration and the ground state becomes a simple translation-invariant state: a ferromagnet (FM) for $J_z > J$, and an in-plane ferroelectric (FE) with a non-zero value of the electrical polarization $\mathbf{P} = \hat{z} \times \langle \bm{\sigma} \rangle$ for $J_z < J$.
On the other hand, at large $J_+$, we expect a configuration that breaks both translation and rotation ($C_{3z}$) symmetries: the phase structure of the bond-dependent term favors different ordering directions around each site, preferring an antiferroelectric (AFE) stripe order with a two-site unit cell. 

To put this intuition on a firmer footing, we carry out a Luttinger-Tisza (LT) analysis of $H$~\cite{luttingerTheoryDipoleInteraction1946,lyons1960method,kaplan2007spin}.
The preferred ordering patterns are determined by the minimum-energy eigenvector of the Fourier-transformed $3 \times 3$ interaction matrix $\mathcal{J}_{\sigma \sigma^\prime}(\q)$ in pseudospin space~\cite{SM}.
Within the weak LT approximation, the pseudospins are only constrained to have unit length on average.
Once such an energy minimum is found, it corresponds to a classical configuration only if it survives the imposition of the strong LT constraint, which dictates that each individual pseudospin must have unit length.

The results from this LT computation are consistent with the expectations outlined previously. 
For small $J_+$, the minimum of $\mathcal{J}_{\sigma \sigma^\prime}(\q)$ lies at the $\Gamma$ point ($\q = 0$), indicating uniform polarization which is either in-plane (ferroelectricity) or out-of-plane (ferromagnetism).
However, beyond a critical value of $J_+$, the minimum of the interaction matrix moves to the $M$ point, indicating the stabilization of an antiferroelectric stripe phase.  
Classically, therefore, the phase diagram of $H$ consists of three simple collinear phases --- this result is further confirmed by a numerical optimization of energy over classical configurations in larger unit cells. 

The LT analysis already offers some insight into the importance of quantum fluctuations in the phase diagram~\cite{KV2014}.
Notably, at the boundary between the FE and stripe phases ($J_+=J$), the energy minima have two distinct features.
First, they are not unique: solutions to the weak LT conditions are satisfied along the lines that connect the $\Gamma$ and the $M$ points, corresponding to collinear incommensurate stripes that interpolate smoothly between the FE and the AFE stripe~\cite{SM}.
This suggests a large degree of degeneracy.
Second, these solutions cannot satisfy the strong LT condition, indicating that they do not admit a simple classical spin configuration.

\textit{Quantum phase diagram.---}
To find the quantum ground state of $H$, we employ large-scale tensor network computations with VUMPS on cylinders of width 4--8 \cite{zauner-stauber_variational_2018,vandammeMPSKit2025}.
We find that quantum fluctuations significantly modify the classical phase diagram over large swaths of parameter space, namely by restricting the orientation of the FE state and inducing a chiral ferromagnet.

For the FE phase, while the classical LT analysis allows for $\langle \bm{\sigma} \rangle$ to point in any direction in the plane, VUMPS ground state optimization finds that the ordering is locked perpendicular to the direction of the lattice vectors.
This may be intuitively understood via the order-by-disorder mechanism~\cite{mcclarty2014order,rousochatzakis2015phase,bergman2007order}---the harmonic zero-point energy of the FE phase, as described by linear spin-wave theory, is lowest when the polarization aligns perpendicular to the lattice directions, as the magnon excitation spectrum is only $C_{3v}$ symmetric~\cite{SM}.

\begin{figure}
    \centering
    \includegraphics{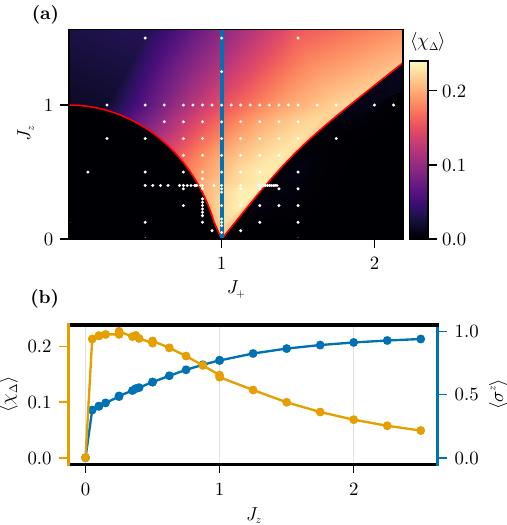}
    \caption{\textbf{(a)} Heatmap of the three-spin scalar chirality $\ev{\chi_\Delta}$ from VUMPS. Quantum phase boundaries are shown in dotted red lines. Data was taken at the white points.
    \textbf{(b)} As $J_z$ decreases along the blue line in (a) [$J_+ = 1$], quantum fluctuations become more significant.
    This is manifested by the decrease in magnetization $\ev{\sigma^z}$ and increase in scalar chirality $\ev{\chi_\Delta}$.
    }
    \label{fig:chiral_fm}
\end{figure}

The most remarkable effect of quantum fluctuations, however, shows up at the phase boundary between the FE and the stripe: the  entire vicinity of the classical phase boundary is replaced by the FM, even as $J_z$ goes to zero~[Fig.~\ref{fig:chiral_fm}(a)]. 
However, this FM state is very distinct from the collinear FM at large $J_z$, which is (nearly) a product state with almost saturated magnetization $\langle \sigma^z_i \rangle \approx 1$. 
By contrast, the FM state at small $J_z$ exhibits strong quantum fluctuations: the size of the ordered moment is quite small~[Fig.~\ref{fig:chiral_fm}(b)].
More significantly, its scalar orbital chirality, $\langle \chi_{ijk} \rangle = \langle \bm{\sigma}_i \cdot (\bm{\sigma}_j \times \bm{\sigma}_k) \rangle$, is non-zero for each triangular plaquette with vertices $\{i,j,k\}$~[Fig.~\ref{fig:chiral_fm}(b)].

\begin{figure}
\centering
\includegraphics[scale=1.0]{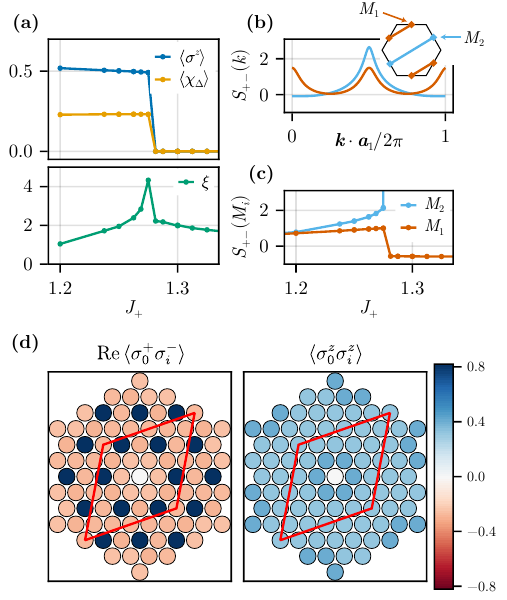}
    \caption{{\textbf{(a)}} (Upper) When we vary $J_+$ at fixed $J_z = 0.4$, we see that at $J_+ = 1.28$, the magnetization $\ev{\sigma^z}$ and chirality $\ev {\chi_\Delta}$ vanish simultaneously.
    (Lower) The correlation length $\xi$ obtained from VUMPS peaks, consistent with a divergence at the transition for the true ground states. This indicates that the transition from FM to stripe is continuous.
    {\textbf{(b)}} On the FM side of the transition, at $J_+ = 1.24$, we observe the onset of $C_{3z}$-symmetry breaking as measured by comparing the structure factor $S_{+-}(k) = \ev{\sigma^+_k \sigma^-_{-k}} - \ev{\sigma_{i=0}^+ \sigma_{i=0}^-}$ at two different $M$ points of the Brillouin zone that are related by threefold rotation. 
    The full structure factor for two cuts in the Brillouin zone at the point $(J_+, J_z) = (1.55, 0.4)$ at circumference $L_y = 6$ and bond dimension $D = 2048$.
    {\textbf{(c)}} As one approaches the stripe phase, this $C_{3z}$ symmetry breaking grows [$J_z=0.4$].
    {\textbf{(d)}} ED computation of the two-point correlation function $\ev{\sigma_0^+\sigma_i^-}$ with the central pseudospin ($0$, white) on a 28-site cluster with periodic boundary conditions (indicated by red rhombus) at $(J_+,J_z)=(1.25,0.2)$. 
    The NN antiferroelectric correlations in plane indicates translation symmetry breaking with  a 4-site unit cell, while the system remains ferromagnetic [$\ev{\sigma_0^z\sigma_i^z} > 0$].}
    \label{fig3}
\end{figure}

To understand the origin of this chirality, let us start with the fully polarized FM: $\ket{\psi_{\rm FM}} = \prod_{i}\otimes\ket{\downarrow}$.
In addition to permuting sites, the action of the $C_{3z}$ rotation attaches a phase $\omega$ to up-spins and $\omega^2$ to down-spins.
Assuming the number of sites is a multiple of $3$, then $\ket{\psi_{\rm FM}}$ is an eigenstate of $C_{3z}$ with eigenvalue 1. 
We may then evolve the state with our Hamiltonian $H$ in imaginary time to reach the actual ground state. 
At each step, the $J_+$ term creates two spin flips, so the spin contribution to the $C_{3z}$ eigenvalue is $\omega^2$. 
To remain a trivial representation of $C_{3z}$, the spatial part of the wavefunction needs to acquire a phase $\omega$ under rotation, which is imparted by the associated phase factors $\gamma_{ij}$.
This intuition is confirmed via an explicit computation (see End Matter)---the $J_+$ term serves to smoothly generate chirality at the expense of $z$-magnetization, altering the character of the ferromagnet while preserving its symmetry properties.

Notably, this mechanism for generating scalar chirality is distinct from previous studies focusing on chiral antiferromagnets~\cite{Esaki2025,park2026spin,Esaki2026,Kim2026}.
There, either an explicit DM term in the Hamiltonian or spin-spin interactions drive a non-zero expectation value of the vector chirality $\bm{D}_{jk} = \langle \bm{\sigma}_j \times \bm{\sigma}_k \rangle$, which combined with a non-zero magnetization $\m_i$ induces a finite, three-body scalar chirality $\langle \chi_{ijk} \rangle \approx \m_ i \cdot \bm{D}_{jk}$.
In our calculation, $|\langle \bm{D}_{jk} \rangle| \lesssim 10^{-7}$, highlighting that the observed scalar chirality arises from correlated quantum fluctuations of different in- and out-of-plane pseudospin components.

{\textit{Phase transitions and the intermediate chiral stripe.---}}
We now characterize the phase transitions present in the $J_+$-$J_z$ phase diagram.
The transition between the FE and the FM is discontinuous (i.e. first-order) except at the special $\mathrm{SU}(2)$-symmetric point $J_z = J$ and $J_+ = 0$.
We see this with VUMPS as a non-differentiable kink in the energy density and a finite region of ``metastability'' around the transition where VUMPS can converge on either FE or FM states, only one of which is the true ground state.
This is expected; the FE and FM phases break different symmetries ($C_{3z}$ and $\cal{T}$ respectively) and a continuous phase transition is thus Landau-forbidden unless fine-tuned.
By the same argument, the transition between the FM (which breaks $\cal{T}$ and $M_y$) and the stripe antiferroelectric (which breaks translation and $C_{3z}$) should also be first order. 
However, the obtained energies vary smoothly and the transfer matrix correlation length appears to diverge~[Fig.~\ref{fig3}(a)], suggesting that the transition is continuous, or at least, weakly first-order.

One possible explanation of these observations is that the system exhibits not a single direct transition, but rather two closely spaced continuous phase transitions: first from the FM to a chiral stripe and then to the collinear stripe phase. 
However, the chiral stripe phase might be difficult to distinguish in VUMPS.
The infinite cylinder geometry explicitly breaks $C_{3z}$, allowing the system to smoothly connect between the FM and the stripe phase.
Indeed, already within the FM region, we observe an asymmetry in the structure factor for the three $C_{3z}$-related $M$ points which grows as one approaches the stripe phase [Fig.~\ref{fig3}(b)].

Given these limitations, we perform an exact diagonalization (ED) study~\cite{xdiag1, xdiag2} near the phase boundary on clusters that explicitly preserve $C_{3z}$.
Between the FM and stripe phase, the system displays two degenerate ground states of opposite chirality, featuring out-of-plane uniform ferromagnetic correlations (which break $\mathcal{T}$) as well as in-plane ferroelectric correlations at the three $M$ points (suggesting translation symmetry breaking)---this indicates the presence of an intermediate stripe phase with finite scalar orbital chirality [Fig.~\ref{fig3}(c)].

We now complement this finite-sized analysis with a thermodynamic mean-field theory of the chiral stripe phase.
We construct a Landau free-energy functional that relaxes the unit-length constraint, enabling us to incorporate fluctuations while being able to accurately describe the vicinity of the continuous phase transition.
The free energy is described by a real soft field $m_0$ and three complex soft fields $\psi_{n}$~\cite{Jin2025}:
\begin{equation}
 \quad \langle \sigma^z_i \rangle = m_0 \quad \text{and} \quad  \langle \sigma^+_{i} \rangle = \sum_{n=1}^{3} \psi_n \,  e^{i \M_n \cdot \r_i}.
 \label{eq:softfields}
\end{equation}
In the End Matter, we show that the chiral stripe phase ($m_0\neq 0$ and $\psi_1 = \omega^2 \psi_2 = \omega  \psi_3 \neq0$, sometimes described as a dense skyrmion crystal~\cite{Okubo2012}) state can be stabilized over both the collinear stripe ($m_0=0$ and $\psi_1\neq 0, \psi_{i>1}=0$) and canted umbrella ($m_0\neq0$ and $\psi_1 = i \, \psi_2\neq0, \psi_{3}=0$) phases~\cite{Kim2026}.
This occurs when the self-repulsion between the amplitudes of the soft fields at the three $M$ points is larger than their mutual repulsion, preferring an equal distribution of the amplitude $|\psi_n|$ over the three $M$ points.

\begin{figure}
    \centering
    \includegraphics[width=1.0\linewidth]{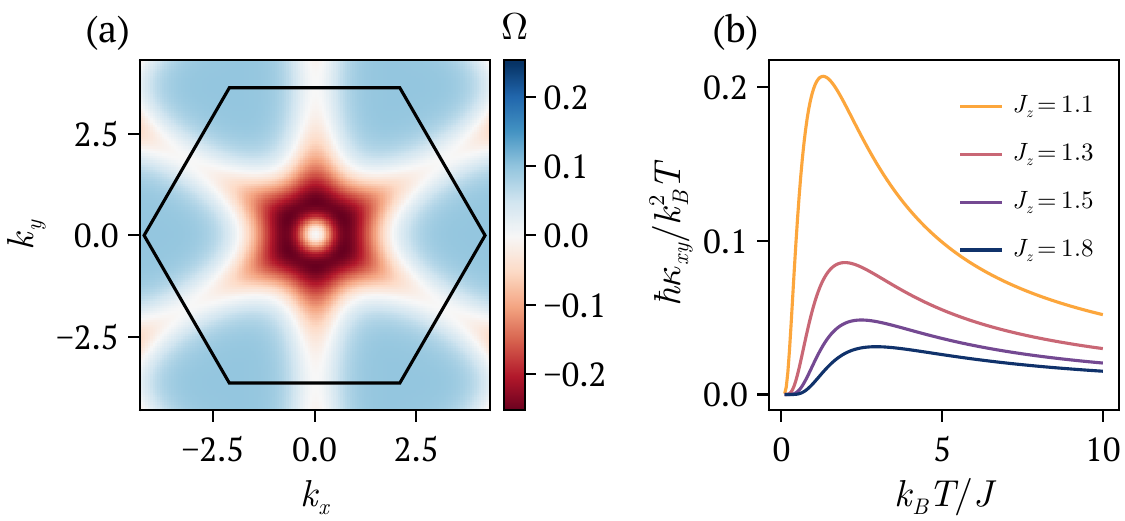}
    \caption{
    Magnon Berry curvature in the FM and the associated thermal Hall conductivity. 
    {\textbf{(a)}} The Berry curvature distribution across the BZ (delineated by the hexagon) in the FM at couplings $J_z = 1.3$ and $J_+ = 1.0$. 
    {\textbf{(b)}} The thermal Hall conductivity along the line $J_+ = 1.0$ for different values of $J_z$ where we see the magnitude of the conductivity increases as the classical transition is approached. }
    \label{fig:FMTHE}
\end{figure}

\textit{Anomalous thermal Hall effect.---}
A key experimental feature of states with spontaneous orbital chirality is an anomalous Hall signal. 
Accordingly, the magnons in the chiral ferromagnet experience the Berry curvature of their Bloch bands, and can therefore exhibit a thermal Hall effect~\cite{katsura2010thermal}. 
To explicitly demonstrate this transport signature, we compute the thermal Hall conductivity $\kappa_{xy}$ for the ferromagnet. 

Starting with the fully polarized FM, we compute the magnon spectra $\varepsilon(\k)$ and wavefunctions using the Holstein-Primakoff transformation~\cite{Auerbach_Book}.
The complex phases in the quadratic magnon Hamiltonian, detailed in the Supplemental Materials (SM)~\cite{SM}, break both time-reversal and reflection $M_y$, inducing Berry curvature $\Omega(\k)$ for the magnon band~[Fig.~\ref{fig:FMTHE}(a)].
Accordingly, the 2D magnon thermal Hall conductivity takes the form, 
\begin{equation}
    \kappa_{xy} = -\frac{k_B^2 T}{\hbar} \int_{\mathrm{BZ}} \frac{d^2 \k}{(2\pi)^2} \left[c_2[n_B(\varepsilon(\k))] - \frac{\pi^2}{3}\right] \Omega(\k),
\end{equation}
where $n_B(\varepsilon(\k)) = [\exp(\varepsilon(\k)/k_B T) - 1]^{-1}$ and $c_2[x] = \int_0^x dt \ln^2\left(\frac{1+t}{t}\right)$~\cite{katsura2010thermal}.
As is evident from Fig.~\ref{fig:FMTHE}(b), the increased magnon thermal Hall conductance from decreasing $J_z/J$ signals the significant enhancement of the orbital chirality. 
We carefully note, however, that spin-wave theory can only partially capture such an effect, because it does not precisely capture the ground state and its excitations, e.g., the enhanced entanglement at smaller $J_z$ as indicated by the VUMPS entanglement spectrum~\cite{SM}.
A complete and accurate characterization of the thermal Hall signal in this entangled chiral ferromagnet is an interesting direction for future study.

\textit{Experimental realizations.---}
Having discussed the wealth of quantum phases observable in a ferromagnetic XXZ Hamiltonian supplemented with a single, symmetry-constrained, magnetization non-conserving coupling, we now turn to discuss settings where our proposed Hamiltonian can be naturally realized. 
We propose three distinct platforms.  

One setting of particular interest is a lattice of moir\'e Wigner molecules in twisted $\Gamma$-valley TMDs~\cite{Reddy2023,Yannouleas,Luo2023}, observed recently in twisted homo-bilayer WS$_2$~\cite{LiWM2024}. 
Here, each molecule is localized at a triangular moir\'e superlattice site and possesses a four-dimensional low-energy state space composed of orbital angular momentum $\sigma^z = \pm 1$ and spin $\pm 1/2$~\cite{KVMC2026}.
Applying a magnetic field normal to the mirror plane does not couple linearly to the orbital pseudospin while polarizing the spins. 
This leads to the Hamiltonian in Eq.~\eqref{eq:H}.
For twisted WS$_2$, this can be achieved with a twist angle of $4.5^\circ$ and a substrate dielectric constant $\varepsilon_r = 11$~\cite{KVMC2026}.

A second setting of contemporary interest is a triangular Rydberg array, where the pseudospins are encoded in two Rydberg states~\cite{adams2019rydberg,browaeys2020many,Bluvstein2021}. 
The dipolar coupling between the Rydberg states naturally includes a spin non-conserving term with the necessary phase structure~\cite{Lienhard2020}. 
While such terms are typically neglected due to level's large energy difference~\cite{tian2025engineering}, applying a magnetic field to ensure they are resonant ensures that the relevant interactions match Eq.~\eqref{eq:H}~\cite{zhao2023floquet}.
Exploring the ferromagnetic phase diagram using the native antiferromagnetic interactions can be done by initializing the system in the highest excited state, and adiabatically preparing the highest excited state of $-H$, or equivalently, the ground state of $H$~\cite{Chen2023}.
We note that, in such platforms, one can explore how long-range interactions may suppress or enhance the system's quantum fluctuations~\cite{Machado2026,Bintz2024,Bintz2026}.

Finally, materials featuring non-Kramers spins (e.g., rare earth compounds on a triangular lattice) naturally exhibit the interactions studied in this work~\cite{Liu2018,Iaconis2018spin,Maksimov2019}.
Whether the signs of the couplings are ferro- or anti-ferromagnetic depends on the microscopic exchange pathways and requires additional exploration.

\textit{Conclusions and outlook.---} 
In this paper, we demonstrated that quantum fluctuations aided by spin-orbit coupling can spontaneously induce scalar orbital chirality in a ferromagnet. 
The orbital chirality arises spontaneously via strong quantum fluctuations near the phase boundary between classically collinear phases, in contrast to previous works~\cite{Kim2026,Esaki2026} where it arises from a non-zero expectation value of time-reversal symmetric vector chirality in combination with an external field-induced magnetization.
This opens up a new category of quantum correlated ferromagnets to explore.

By showing that ferromagnetic Hamiltonians can exhibit strong quantum fluctuations when perturbed with bond-dependent couplings, our work naturally motivates several future directions. 
First, we studied the spontaneous emergence of chirality in the $T = 0$ phase diagram, but not the effect of thermal fluctuations, which have been proposed to enhance such chirality within the framework of spin-wave theory in chiral antiferromagnets~\cite{Esaki2025}. 
Second, the chiral orbital ferromagnet has low-energy magnon excitations, but their quantum wavefunctions are expected to be very distinct from those of conventional spin-wave excitations in a product state FM. 
It remains an open question how orbital chirality ``dresses'' the magnon wavefunction and impacts the magnon spectrum and dynamics. 
Finally, such phase-constrained anisotropic perturbations to an XXZ Hamiltonian are not specific to the triangular lattice: they will realistically be present in any lattice with local pseudospin degrees of freedom encoded in degenerate chiral/anti-chiral orbitals.  
We leave a detailed construction of such models and a study of their quantum phase diagrams to future work. 

\textit{Acknowledgements.---} It is a pleasure to thank Jason Alicea, Marcus Bintz, Stefan Divic and Joel Moore for valuable conversations. 
The computations in this work were done using Bridges-2 at PSC through allocation PHY250361 from the Advanced Cyberinfrastructure Coordination Ecosystem: Services \& Support (ACCESS) program~\cite{access}, which is supported by U.S. National Science Foundation grants No. 2138259, No. 2138286, No. 2138307, No. 2137603, and No. 2138296.
F.M.~was supported by the Netherlands Organisation for Scientific Research (NWO/OCW), as part of Quantum Limits (project
number SUMMIT.1.1016).

\bibliography{orbital}

@article{cincioCharacterizingTopologicalOrder2013,
   title={Characterizing Topological Order by Studying the Ground States on an Infinite Cylinder},
   volume={110},
   ISSN={1079-7114},
   url={http://dx.doi.org/10.1103/PhysRevLett.110.067208},
   DOI={10.1103/physrevlett.110.067208},
   number={6},
   journal={Physical Review Letters},
   publisher={American Physical Society (APS)},
   author={Cincio, L. and Vidal, G.},
   year={2013},
   month=Feb }

@article{vanderstraeten2019,
   title={Tangent-space methods for uniform matrix product states},
   ISSN={2590-1990},
   url={http://dx.doi.org/10.21468/SciPostPhysLectNotes.7},
   DOI={10.21468/scipostphyslectnotes.7},
   journal={SciPost Physics Lecture Notes},
   publisher={Stichting SciPost},
   author={Vanderstraeten, Laurens and Haegeman, Jutho and Verstraete, Frank},
   year={2019},
   month=Jan }

@article{ciracMatrixProductStates2021,
   title={Matrix product states and projected entangled pair states: Concepts, symmetries, theorems},
   volume={93},
   ISSN={1539-0756},
   url={http://dx.doi.org/10.1103/RevModPhys.93.045003},
   DOI={10.1103/revmodphys.93.045003},
   number={4},
   journal={Reviews of Modern Physics},
   publisher={American Physical Society (APS)},
   author={Cirac, J. Ignacio and Pérez-García, David and Schuch, Norbert and Verstraete, Frank},
   year={2021},
   month=Dec }

@article{Lienhard2020,
  title = {Realization of a Density-Dependent Peierls Phase in a Synthetic, Spin-Orbit Coupled Rydberg System},
  author = {Lienhard, Vincent and Scholl, Pascal and Weber, Sebastian and Barredo, Daniel and de L\'es\'eleuc, Sylvain and Bai, Rukmani and Lang, Nicolai and Fleischhauer, Michael and B\"uchler, Hans Peter and Lahaye, Thierry and Browaeys, Antoine},
  journal = {Phys. Rev. X},
  volume = {10},
  issue = {2},
  pages = {021031},
  numpages = {11},
  year = {2020},
  month = {May},
  publisher = {American Physical Society},
  doi = {10.1103/PhysRevX.10.021031},
  url = {https://link.aps.org/doi/10.1103/PhysRevX.10.021031}
}

@ARTICLE{Chen2023,
       author = {{Chen}, Cheng and {Bornet}, Guillaume and {Bintz}, Marcus and {Emperauger}, Gabriel and {Leclerc}, Lucas and {Liu}, Vincent S. and {Scholl}, Pascal and {Barredo}, Daniel and {Hauschild}, Johannes and {Chatterjee}, Shubhayu and {Schuler}, Michael and {L{\"a}uchli}, Andreas M. and {Zaletel}, Michael P. and {Lahaye}, Thierry and {Yao}, Norman Y. and {Browaeys}, Antoine},
        title = "{Continuous symmetry breaking in a two-dimensional Rydberg array}",
      journal = {\nat},
         year = 2023,
        month = apr,
       volume = {616},
       number = {7958},
        pages = {691-695},
          doi = {10.1038/s41586-023-05859-2},
archivePrefix = {arXiv},
       eprint = {2207.12930},
 primaryClass = {cond-mat.quant-gas},
       adsurl = {https://ui.adsabs.harvard.edu/abs/2023Natur.616..691C}
}

@Article{KVMC2026,
	title={{Spin-orbital magnetism in moiré Wigner molecules}},
	author={Ahmed Khalifa and Rokas Veitas and Francisco Machado and Shubhayu Chatterjee},
	journal={SciPost Phys.},
	volume={20},
	pages={090},
	year={2026},
	publisher={SciPost},
	doi={10.21468/SciPostPhys.20.3.090},
	url={https://scipost.org/10.21468/SciPostPhys.20.3.090},
}

@article{luttingerTheoryDipoleInteraction1946,
  title = {Theory of {{Dipole Interaction}} in {{Crystals}}},
  author = {Luttinger, J. M. and Tisza, L.},
  year = 1946,
  month = dec,
  journal = {Physical Review},
  volume = {70},
  number = {11-12},
  pages = {954--964},
  issn = {0031-899X},
  doi = {10.1103/PhysRev.70.954},
  urldate = {2026-03-26},
  copyright = {http://link.aps.org/licenses/aps-default-license},
  langid = {english}
}

@article{lyons1960method,
  title={Method for determining ground-state spin configurations},
  author={Lyons, DH and Kaplan, TA},
  journal={Physical Review},
  volume={120},
  number={5},
  pages={1580},
  year={1960},
  publisher={APS}
}

@article{kaplan2007spin,
  title={Spin ordering in three-dimensional crystals with strong competing exchange interactions},
  author={Kaplan, TA and Menyuk, N},
  journal={Philosophical Magazine},
  volume={87},
  number={25},
  pages={3711--3785},
  year={2007},
  publisher={Taylor \& Francis}
}

@ARTICLE{KV2014,
       author = {{Kimchi}, Itamar and {Vishwanath}, Ashvin},
        title = "{Kitaev-Heisenberg models for iridates on the triangular, hyperkagome, kagome, fcc, and pyrochlore lattices}",
      journal = {\prb},
         year = 2014,
        month = jan,
       volume = {89},
       number = {1},
          eid = {014414},
        pages = {014414},
          doi = {10.1103/PhysRevB.89.014414},
archivePrefix = {arXiv},
       eprint = {1303.3290},
 primaryClass = {cond-mat.str-el},
       adsurl = {https://ui.adsabs.harvard.edu/abs/2014PhRvB..89a4414K}
}

@article{Reddy2023,
  title = {Artificial Atoms, Wigner Molecules, and an Emergent Kagome Lattice in Semiconductor Moir\'e Superlattices},
  author = {Reddy, Aidan P. and Devakul, Trithep and Fu, Liang},
  journal = {Phys. Rev. Lett.},
  volume = {131},
  issue = {24},
  pages = {246501},
  numpages = {6},
  year = {2023},
  month = {Dec},
  publisher = {American Physical Society},
  doi = {10.1103/PhysRevLett.131.246501},
  url = {https://link.aps.org/doi/10.1103/PhysRevLett.131.246501}
}

@article{Yannouleas,
  title = {Quantum Wigner molecules in moir\'e materials},
  author = {Yannouleas, Constantine and Landman, Uzi},
  journal = {Phys. Rev. B},
  volume = {108},
  issue = {12},
  pages = {L121411},
  numpages = {7},
  year = {2023},
  month = {Sep},
  publisher = {American Physical Society},
  doi = {10.1103/PhysRevB.108.L121411},
  url = {https://link.aps.org/doi/10.1103/PhysRevB.108.L121411}
}

@ARTICLE{Luo2023,
       author = {{Luo}, Di and {Reddy}, Aidan P. and {Devakul}, Trithep and {Fu}, Liang},
        title = "{Artificial intelligence for artificial materials: moir{\'e} atom}",
      journal = {arXiv e-prints},
         year = 2023,
        month = mar,
          eid = {arXiv:2303.08162},
          doi = {10.48550/arXiv.2303.08162},
archivePrefix = {arXiv},
       eprint = {2303.08162},
 primaryClass = {cond-mat.str-el},
       adsurl = {https://ui.adsabs.harvard.edu/abs/2023arXiv230308162L}
}

@article{LiWM2024,
author = {Hongyuan Li  and Ziyu Xiang  and Aidan P. Reddy  and Trithep Devakul  and Renee Sailus  and Rounak Banerjee  and Takashi Taniguchi  and Kenji Watanabe  and Sefaattin Tongay  and Alex Zettl  and Liang Fu  and Michael F. Crommie  and Feng Wang },
title = {Wigner molecular crystals from multielectron moiré artificial atoms},
journal = {Science},
volume = {385},
number = {6704},
pages = {86-91},
year = {2024},
doi = {10.1126/science.adk1348},
URL = {https://www.science.org/doi/abs/10.1126/science.adk1348}
}

@article{adams2019rydberg,
  title={Rydberg atom quantum technologies},
  author={Adams, Charles S and Pritchard, Jonathan D and Shaffer, James P},
  journal={Journal of Physics B: Atomic, Molecular and Optical Physics},
  volume={53},
  number={1},
  pages={012002},
  year={2019},
  publisher={IOP Publishing}
}

@article{browaeys2020many,
  title={Many-body physics with individually controlled Rydberg atoms},
  author={Browaeys, Antoine and Lahaye, Thierry},
  journal={Nature Physics},
  volume={16},
  number={2},
  pages={132--142},
  year={2020},
  publisher={Nature Publishing Group UK London}
}

@article{Bluvstein2021,
  title={Controlling quantum many-body dynamics in driven Rydberg atom arrays},
  author={Bluvstein, Dolev and Omran, Ahmed and Levine, Harry and Keesling, Alexander and Semeghini, Giulia and Ebadi, Sepehr and Wang, Tout T and Michailidis, Alexios A and Maskara, Nishad and Ho, Wen Wei and others},
  journal={Science},
  volume={371},
  number={6536},
  pages={1355--1359},
  year={2021},
  publisher={American Association for the Advancement of Science}
}

@article{tian2025engineering,
  title={Engineering frustrated Rydberg spin models by graphical Floquet modulation},
  author={Tian, Mingsheng and Samajdar, Rhine and Gadway, Bryce},
  journal={Physical review letters},
  volume={135},
  number={25},
  pages={253001},
  year={2025},
  publisher={APS}
}

@article{zhao2023floquet,
  title={Floquet-tailored Rydberg interactions},
  author={Zhao, Luheng and Lee, Michael Dao Kang and Aliyu, Mohammad Mujahid and Loh, Huanqian},
  journal={Nature Communications},
  volume={14},
  number={1},
  pages={7128},
  year={2023},
  publisher={Nature Publishing Group UK London}
}

@ARTICLE{Liu2018,
       author = {{Liu}, Changle and {Li}, Yao-Dong and {Chen}, Gang},
        title = "{Selective measurements of intertwined multipolar orders: Non-Kramers doublets on a triangular lattice}",
      journal = {\prb},
         year = 2018,
        month = jul,
       volume = {98},
       number = {4},
          eid = {045119},
        pages = {045119},
          doi = {10.1103/PhysRevB.98.045119},
archivePrefix = {arXiv},
       eprint = {1805.01865},
 primaryClass = {cond-mat.str-el},
       adsurl = {https://ui.adsabs.harvard.edu/abs/2018PhRvB..98d5119L}
}

@inproceedings{access,
author = {Boerner, Timothy J. and Deems, Stephen and Furlani, Thomas R. and Knuth, Shelley L. and Towns, John},
title = {ACCESS: Advancing Innovation: NSF’s Advanced Cyberinfrastructure Coordination Ecosystem: Services \& Support},
year = {2023},
isbn = {9781450399852},
publisher = {Association for Computing Machinery},
address = {New York, NY, USA},
url = {https://doi.org/10.1145/3569951.3597559},
doi = {10.1145/3569951.3597559},
pages = {173–176},
numpages = {4},
location = {Portland, OR, USA},
series = {PEARC '23}
}

@article{zauner-stauber_variational_2018,
    title = {Variational optimization algorithms for uniform matrix product states},
    volume = {97},
    issn = {2469-9950, 2469-9969},
    url = {http://arxiv.org/abs/1701.07035},
    doi = {10.1103/PhysRevB.97.045145},
    number = {4},
    urldate = {2024-04-05},
    journal = {Physical Review B},
    author = {Zauner-Stauber, V. and Vanderstraeten, L. and Fishman, M. T. and Verstraete, F. and Haegeman, J.},
    month = jan,
    year = {2018},
    note = {arXiv:1701.07035 [cond-mat, physics:quant-ph]},
    pages = {045145},
}

@software{vandammeMPSKit2025,
  title = {{{MPSKit}}},
  author = {Van Damme, Maarten and Devos, Lukas and Haegeman, Jutho},
  year = {2025},
  date = {2025-09-05},
  doi = {10.5281/ZENODO.10654900},
  url = {https://zenodo.org/doi/10.5281/zenodo.10654900},
  urldate = {2025-09-24},
  organization = {Zenodo},
  version = {v0.13.5}
}

@ARTICLE{katsura2010thermal,
       author = {{Katsura}, Hosho and {Nagaosa}, Naoto and {Lee}, Patrick A.},
        title = "{Theory of the Thermal Hall Effect in Quantum Magnets}",
      journal = {\prl},
         year = 2010,
        month = feb,
       volume = {104},
       number = {6},
          eid = {066403},
        pages = {066403},
          doi = {10.1103/PhysRevLett.104.066403},
archivePrefix = {arXiv},
       eprint = {0904.3427},
 primaryClass = {cond-mat.str-el},
       adsurl = {https://ui.adsabs.harvard.edu/abs/2010PhRvL.104f6403K}
}

@ARTICLE{Iaconis2018spin,
       author = {{Iaconis}, Jason and {Liu}, Chunxiao and {Hal{\'a}sz}, G{\'a}bor and {Balents}, Leon},
        title = "{Spin Liquid versus Spin Orbit Coupling on the Triangular Lattice}",
      journal = {SciPost Physics},
         year = 2018,
        month = jan,
       volume = {4},
       number = {1},
          eid = {003},
        pages = {003},
          doi = {10.21468/SciPostPhys.4.1.003},
archivePrefix = {arXiv},
       eprint = {1708.07856},
 primaryClass = {cond-mat.str-el},
       adsurl = {https://ui.adsabs.harvard.edu/abs/2018ScPP....4....3I}
}

@ARTICLE{Maksimov2019,
       author = {{Maksimov}, P.~A. and {Zhu}, Zhenyue and {White}, Steven R. and {Chernyshev}, A.~L.},
        title = "{Anisotropic-Exchange Magnets on a Triangular Lattice: Spin Waves, Accidental Degeneracies, and Dual Spin Liquids}",
      journal = {Physical Review X},
         year = 2019,
        month = apr,
       volume = {9},
       number = {2},
          eid = {021017},
        pages = {021017},
          doi = {10.1103/PhysRevX.9.021017},
archivePrefix = {arXiv},
       eprint = {1811.05983},
 primaryClass = {cond-mat.str-el},
       adsurl = {https://ui.adsabs.harvard.edu/abs/2019PhRvX...9b1017M}
}

@book{Auerbach_Book,
  title={Interacting Electrons and Quantum Magnetism},
  author={Auerbach, Assa},
  year={1998},
  publisher={Springer Science \& Business Media}
}

@article{Savary2016QSL,
  title={Quantum spin liquids: a review},
  author={Savary, Lucile and Balents, Leon},
  journal={Reports on Progress in Physics},
  volume={80},
  number={1},
  pages={016502},
  year={2016},
  publisher={IOP Publishing}
}

@article{Zhou2017QSL,
  title={Quantum spin liquid states},
  author={Zhou, Yi and Kanoda, Kazushi and Ng, Tai-Kai},
  journal={Reviews of Modern Physics},
  volume={89},
  number={2},
  pages={025003},
  year={2017},
  publisher={APS}
}

@article{Broholm2020QSL,
  title={Quantum spin liquids},
  author={Broholm, C and Cava, RJ and Kivelson, SA and Nocera, DG and Norman, MR and Senthil, T},
  journal={Science},
  volume={367},
  number={6475},
  pages={eaay0668},
  year={2020},
  publisher={American Association for the Advancement of Science}
}

@article{Clark2021QSL,
  title={Quantum spin liquids from a materials perspective},
  author={Clark, Lucy and Abdeldaim, Aly H},
  journal={Annual Review of Materials Research},
  volume={51},
  pages={495--519},
  year={2021},
  publisher={Annual Reviews}
}

@article{senthil2004deconfined,
  title={Deconfined quantum critical points},
  author={Senthil, Todadri and Vishwanath, Ashvin and Balents, Leon and Sachdev, Subir and Fisher, Matthew PA},
  journal={Science},
  volume={303},
  number={5663},
  pages={1490--1494},
  year={2004},
  publisher={American Association for the Advancement of Science}
}

@ARTICLE{ChongWangDQCP2017,
       author = {{Wang}, Chong and {Nahum}, Adam and {Metlitski}, Max A. and {Xu}, Cenke and {Senthil}, T.},
        title = "{Deconfined Quantum Critical Points: Symmetries and Dualities}",
      journal = {Physical Review X},
         year = 2017,
        month = jul,
       volume = {7},
       number = {3},
          eid = {031051},
        pages = {031051},
          doi = {10.1103/PhysRevX.7.031051},
archivePrefix = {arXiv},
       eprint = {1703.02426},
 primaryClass = {cond-mat.str-el},
       adsurl = {https://ui.adsabs.harvard.edu/abs/2017PhRvX...7c1051W}
}

@article{senthil2024deconfined,
  title={Deconfined quantum critical points: a review},
  author={Senthil, T},
  journal={50 Years of the Renormalization Group: Dedicated to the Memory of Michael E Fisher},
  pages={169--195},
  year={2024},
  publisher={World Scientific}
}

@article{ramirez1994strongly,
  title={Strongly Geometrically Frustrated Magnets},
  author={Ramirez, A.~P.},
  journal={Annual Review of Materials Science},
  volume={24},
  number={1},
  pages={453--480},
  year={1994},
  publisher={Annual Reviews 4139 El Camino Way, PO Box 10139, Palo Alto, CA 94303-0139, USA}
}

@book{lacroix2011introduction,
  title={Introduction to Frustrated Magnetism: Materials, Experiments, Theory},
  author={Lacroix, Claudine and Mendels, Philippe and Mila, Fr\'ed\'eric},
  year={2011},
  publisher={Springer Science \& Business Media}
}

@article{li2008entanglement,
  title = {Entanglement Spectrum as a Generalization of Entanglement Entropy: Identification of Topological Order in Non-Abelian Fractional Quantum Hall Effect States},
  author = {Li, Hui and Haldane, F. D. M.},
  journal = {Phys. Rev. Lett.},
  volume = {101},
  issue = {1},
  pages = {010504},
  numpages = {4},
  year = {2008},
  month = {Jul},
  publisher = {American Physical Society},
  doi = {10.1103/PhysRevLett.101.010504},
  url = {https://link.aps.org/doi/10.1103/PhysRevLett.101.010504}
}

@ARTICLE{Sondhi2008,
       author = {{Castelnovo}, C. and {Moessner}, R. and {Sondhi}, S.~L.},
        title = "{Magnetic monopoles in spin ice}",
      journal = {\nat},
         year = 2008,
        month = jan,
       volume = {451},
       number = {7174},
        pages = {42-45},
          doi = {10.1038/nature06433},
archivePrefix = {arXiv},
       eprint = {0710.5515},
 primaryClass = {cond-mat.str-el},
       adsurl = {https://ui.adsabs.harvard.edu/abs/2008Natur.451...42C}
}

@article{mcclarty2014order,
  title={Order-by-disorder in the XY pyrochlore antiferromagnet},
  author={McClarty, Paul A and Stasiak, Pawel and Gingras, Michel JP},
  journal={Physical Review B},
  volume={89},
  number={2},
  pages={024425},
  year={2014},
  publisher={APS}
}

@article{rousochatzakis2015phase,
  title={Phase diagram and quantum order by disorder in the Kitaev K 1- K 2 honeycomb magnet},
  author={Rousochatzakis, Ioannis and Reuther, Johannes and Thomale, Ronny and Rachel, Stephan and Perkins, Natalia B},
  journal={Physical Review X},
  volume={5},
  number={4},
  pages={041035},
  year={2015},
  publisher={APS}
}

@article{bergman2007order,
  title={Order-by-disorder and spiral spin-liquid in frustrated diamond-lattice antiferromagnets},
  author={Bergman, Doron and Alicea, Jason and Gull, Emanuel and Trebst, Simon and Balents, Leon},
  journal={Nature Physics},
  volume={3},
  number={7},
  pages={487--491},
  year={2007},
  publisher={Nature Publishing Group UK London}
}

@article{park2026spin,
  title={Spin-orbit-induced Instability and Finite-Temperature Stabilization of a Triangular-lattice Supersolid},
  author={Park, Seongjun and Park, Sung-Min and Oh, Yun-Tak and Lee, Hyun-Yong and Moon, Eun-Gook},
  journal={arXiv preprint arXiv:2601.20963},
  year={2026}
}

@ARTICLE{Kim2026,
       author = {{Kim}, Daesik and {Jeon}, Hyojae and {Park}, Seongjun and {Lee}, Seungho and {Han}, Jung Hoon and {Lee}, Hyun-Yong},
        title = "{Spin-orbit-induced quantum chiral phases}",
      journal = {arXiv e-prints},
         year = 2026,
        month = apr,
          eid = {arXiv:2604.19161},
        pages = {arXiv:2604.19161},
          doi = {10.48550/arXiv.2604.19161},
archivePrefix = {arXiv},
       eprint = {2604.19161},
 primaryClass = {cond-mat.str-el},
       adsurl = {https://ui.adsabs.harvard.edu/abs/2026arXiv260419161K}
}

@ARTICLE{Esaki2026,
       author = {{Esaki}, Nanse and {Go}, Gyungchoon and {Kim}, Se Kwon},
        title = "{Quantum Scalar Spin Chirality in Coplanar Kagome Antiferromagnets}",
      journal = {arXiv e-prints},
         year = 2026,
        month = apr,
          eid = {arXiv:2604.27632},
        pages = {arXiv:2604.27632},
          doi = {10.48550/arXiv.2604.27632},
archivePrefix = {arXiv},
       eprint = {2604.27632},
 primaryClass = {cond-mat.mes-hall},
       adsurl = {https://ui.adsabs.harvard.edu/abs/2026arXiv260427632E}
}

@article{Esaki2025,
  title = {Magnon-induced scalar spin chirality in kagome and honeycomb ferromagnets},
  author = {Esaki, Nanse and Go, Gyungchoon and Kim, Se Kwon},
  journal = {Phys. Rev. B},
  volume = {112},
  issue = {10},
  pages = {104440},
  numpages = {10},
  year = {2025},
  month = {Sep},
  publisher = {American Physical Society},
  doi = {10.1103/fhwj-76zp},
  url = {https://link.aps.org/doi/10.1103/fhwj-76zp}
}

@ARTICLE{Okubo2012,
       author = {{Okubo}, Tsuyoshi and {Chung}, Sungki and {Kawamura}, Hikaru},
        title = "{Multiple-q States and the Skyrmion Lattice of the Triangular-Lattice Heisenberg Antiferromagnet under Magnetic Fields}",
      journal = {\prl},
         year = 2012,
        month = jan,
       volume = {108},
       number = {1},
          eid = {017206},
        pages = {017206},
          doi = {10.1103/PhysRevLett.108.017206},
archivePrefix = {arXiv},
       eprint = {1109.6161},
 primaryClass = {cond-mat.stat-mech},
       adsurl = {https://ui.adsabs.harvard.edu/abs/2012PhRvL.108a7206O}
}

@article{SM,
	Journal={See supplementary Materials for additional details.}
}

@book{sachdev:2011,
  title = {Quantum {{Phase Transitions}}},
  author = {Sachdev, Subir},
  year = 2011,
  edition = {2},
  publisher = {Cambridge University Press},
  address = {Cambridge},
  doi = {10.1017/CBO9780511973765},
  isbn = {978-0-521-51468-2}
}

@Article{xdiag1,
	title={{XDiag: Exact diagonalization for quantum many-body systems}},
	author={Alexander Wietek and Luke Staszewski and Martin Ulaga and Paul L. Ebert and Hannes Karlsson and Siddhartha Sarkar and Leyna Shackleton and Aritra Sinha and Rafael D. Soares},
	journal={SciPost Phys. Codebases},
	pages={70},
	year={2026},
	publisher={SciPost},
	doi={10.21468/SciPostPhysCodeb.70},
	url={https://scipost.org/10.21468/SciPostPhysCodeb.70},
}

@Article{xdiag2,
	title={{Codebase release 0.4 for XDiag}},
	author={Alexander Wietek and Luke Staszewski and Martin Ulaga and Paul L. Ebert and Hannes Karlsson and Siddhartha Sarkar and Leyna Shackleton and Aritra Sinha and Rafael D. Soares},
	journal={SciPost Phys. Codebases},
	pages={70-r0.4},
	year={2026},
	publisher={SciPost},
	doi={10.21468/SciPostPhysCodeb.70-r0.4},
	url={https://scipost.org/10.21468/SciPostPhysCodeb.70-r0.4},
}

@article{PhysRevB.94.174424,
  title = {Semiclassical ground-state phase diagram and $\text{multi-}Q$ phase of a spin-orbit-coupled model on triangular lattice},
  author = {Liu, Changle and Yu, Rong and Wang, Xiaoqun},
  journal = {Phys. Rev. B},
  volume = {94},
  issue = {17},
  pages = {174424},
  numpages = {9},
  year = {2016},
  month = {Nov},
  publisher = {American Physical Society},
  doi = {10.1103/PhysRevB.94.174424},
  url = {https://link.aps.org/doi/10.1103/PhysRevB.94.174424}
}

@article{PhysRevLett.115.167203,
  title = {Rare-Earth Triangular Lattice Spin Liquid: A Single-Crystal Study of ${\mathrm{YbMgGaO}}_{4}$},
  author = {Li, Yuesheng and Chen, Gang and Tong, Wei and Pi, Li and Liu, Juanjuan and Yang, Zhaorong and Wang, Xiaoqun and Zhang, Qingming},
  journal = {Phys. Rev. Lett.},
  volume = {115},
  issue = {16},
  pages = {167203},
  numpages = {6},
  year = {2015},
  month = {Oct},
  publisher = {American Physical Society},
  doi = {10.1103/PhysRevLett.115.167203},
  url = {https://link.aps.org/doi/10.1103/PhysRevLett.115.167203}
}

@article{Jin2025,
  title = {Phenomenological Ginzburg-Landau theory for triple-Q magnetic orders on a hexagonal lattice},
  author = {Jin, Jin-Tao and Zhou, Yi},
  journal = {Phys. Rev. B},
  volume = {112},
  issue = {22},
  pages = {224434},
  numpages = {13},
  year = {2025},
  month = {Dec},
  publisher = {American Physical Society},
  doi = {10.1103/nvnm-rtwd},
  url = {https://link.aps.org/doi/10.1103/nvnm-rtwd}
}

@ARTICLE{Bintz2024,
       author = {{Bintz}, Marcus and {Liu}, Vincent S. and {Hauschild}, Johannes and {Khalifa}, Ahmed and {Chatterjee}, Shubhayu and {Zaletel}, Michael P. and {Yao}, Norman Y.},
        title = "{Dirac spin liquid in quantum dipole arrays}",
      journal = {arXiv e-prints},
         year = 2024,
        month = may,
          eid = {arXiv:2406.00098},
        pages = {arXiv:2406.00098},
          doi = {10.48550/arXiv.2406.00098},
archivePrefix = {arXiv},
       eprint = {2406.00098},
 primaryClass = {cond-mat.str-el},
       adsurl = {https://ui.adsabs.harvard.edu/abs/2024arXiv240600098B}
}

@ARTICLE{Machado2026,
       author = {{Machado}, Francisco and {Chern}, Sabrina and {Zaletel}, Michael P. and {Yao}, Norman Y.},
        title = "{A Dipolar Chiral Spin Liquid on the Breathed Kagome Lattice}",
      journal = {arXiv e-prints},
         year = 2026,
        month = mar,
          eid = {arXiv:2603.25784},
        pages = {arXiv:2603.25784},
          doi = {10.48550/arXiv.2603.25784},
archivePrefix = {arXiv},
       eprint = {2603.25784},
 primaryClass = {cond-mat.quant-gas},
       adsurl = {https://ui.adsabs.harvard.edu/abs/2026arXiv260325784M}
}

@ARTICLE{Bintz2026,
       author = {{Bintz}, Marcus and {Khalifa}, Ahmed and {Liu}, Vincent S. and {Hauschild}, Johannes and {Zaletel}, Michael P. and {Chatterjee}, Shubhayu and {Yao}, Norman Y.},
        title = "{Ground states of quantum XY dipoles on the Archimedean lattices}",
      journal = {arXiv e-prints},
         year = 2026,
        month = may,
          eid = {arXiv:2605.07685},
        pages = {arXiv:2605.07685},
archivePrefix = {arXiv},
       eprint = {2605.07685},
 primaryClass = {cond-mat.str-el},
       adsurl = {https://ui.adsabs.harvard.edu/abs/2026arXiv260507685B}
}

@book{fradkin2021quantum,
  title={Quantum Field Theory: An Integrated Approach},
  author={Fradkin, Eduardo},
  year={2021},
  publisher={Princeton University Press}
}

@book{sachdev2023quantum,
  title={Quantum Phases of Matter},
  author={Sachdev, Subir},
  year={2023},
  publisher={Cambridge University Press}
}

@software{kofod2025,
  title = {{{JuliaNLSolvers}}/{{Optim}}.Jl: V1.13.2 (Docs)},
  shorttitle = {{{JuliaNLSolvers}}/{{Optim}}.Jl},
  author = {Patrick Kofod Mogensen and John Myles White and Asbjørn Nilsen Riseth and Tim Holy and Miles Lubin and Christof and Andreas Noack and Antoine Levitt and Benoît Legat and Christoph Ortner and Blake Johnson and Christopher Rackauckas and Yichao Yu and Kristoffer Carlsson and Dahua Lin and Arno Strouwen and Josua Grawitter and Takafumi Arakaki and Benoît Pasquier and {abhro} and Thomas R. Covert and Ron Rock and Michael Creel and {cossio} and Jeffrey Regier and David Widmann and Ben Kuhn and Alexey Stukalov},
  date = {2025-06-12},
  doi = {10.5281/ZENODO.15649599},
  url = {https://zenodo.org/doi/10.5281/zenodo.15649599},
  urldate = {2025-06-27},
  organization = {Zenodo},
  version = {v1.13.2}
}

@article{edelman1998,
  title={The geometry of algorithms with orthogonality constraints},
  author={Edelman, Alan and Arias, Tom{\'a}s A and Smith, Steven T},
  journal={SIAM journal on Matrix Analysis and Applications},
  volume={20},
  number={2},
  pages={303--353},
  year={1998},
  publisher={SIAM}
}

@incollection{absil2008,
  title={Optimization algorithms on matrix manifolds},
  author={Absil, P-A and Mahony, Robert and Sepulchre, Rodolphe},
  booktitle={Optimization Algorithms on Matrix Manifolds},
  year={2009},
  publisher={Princeton University Press}
}

@note{fnED,
note="We chose $N = 28$ for ED as it is the largest feasible cluster that includes all three M points."
}

@Article{tenpy2024,
    title={{Tensor network Python (TeNPy) version 1}},
    author={Johannes Hauschild and Jakob Unfried and Sajant Anand and Bartholomew Andrews and Marcus Bintz and Umberto Borla and Stefan Divic and Markus Drescher and Jan Geiger and Martin Hefel and Kévin Hémery and Wilhelm Kadow and Jack Kemp and Nico Kirchner and Vincent S. Liu and Gunnar Möller and Daniel Parker and Michael Rader and Anton Romen and Samuel Scalet and Leon Schoonderwoerd and Maximilian Schulz and Tomohiro Soejima and Philipp Thoma and Yantao Wu and Philip Zechmann and Ludwig Zweng and Roger S. K. Mong and Michael P. Zaletel and Frank Pollmann},
    journal={SciPost Phys. Codebases},
    pages={41},
    year={2024},
    publisher={SciPost},
    doi={10.21468/SciPostPhysCodeb.41},
    url={https://scipost.org/10.21468/SciPostPhysCodeb.41},
}

@article{szasz2020chiral,
  title={Chiral spin liquid phase of the triangular lattice Hubbard model: a density matrix renormalization group study},
  author={Szasz, Aaron and Motruk, Johannes and Zaletel, Michael P and Moore, Joel E},
  journal={Physical Review X},
  volume={10},
  number={2},
  pages={021042},
  year={2020},
  publisher={APS}
}

\section{END MATTER}

\subsection{Chirality growth under imaginary time evolution}

In the main text, we provided a symmetry-based argument for the generation of orbital chirality $\langle \chi_\Delta \rangle$ when the fully polarized FM is evolved in imaginary time by $H$ to reach its actual ground state --- the chiral FM. 
Here, we complement this argument via a concrete perturbative calculation of $\langle \chi_\Delta \rangle$ in the state $\ket{\psi_\tau} = e^{- \tau H}\ket{\psi_{\rm FM}}$ at short (imaginary) time $\tau$, and a numerically exact evaluation of $\langle \chi_\Delta(\tau) \rangle$ that shows how the orbital chirality rapidly approaches its ground state value. 

For the perturbative computation, it is useful to first build some intuition by considering the action of the magnetization-non-conserving part of $H$ on a fully polarized state of a single triangular plaquette $\Delta_{ijk}$ with vertices $\{ijk\}$ labeled in an anticlockwise fashion:
\begin{equation}
h_{\Delta_{ijk}} \ket{\downarrow \downarrow \downarrow} \equiv  (\sigma^+_i \sigma^+_j + \omega \, \sigma^+_j \sigma^+_k + \omega^2 \sigma^+_k \sigma^+_i) \ket{\downarrow \downarrow \downarrow}.
\end{equation}
Writing the chirality operator $\chi_{\Delta_{ijk}} = \bm{\sigma}_i \cdot (\bm{\sigma}_j \times \bm{\sigma}_k)$ as
\begin{equation}
\chi_{\Delta_{ijk}} = \frac{i}{2}\left[ \sigma^z_i (\sigma^+_j \sigma^-_k - \sigma^-_j \sigma^+_k) + \text{cyclic permutations}\right],
\end{equation}
it is easy to check that $\chi_{\Delta_{ijk}} \ket{\downarrow \downarrow \downarrow} = 0$, while the state $h_{\Delta_{ijk}} \ket{\downarrow \downarrow \downarrow}$ is an eigenstate of $\chi_{\Delta_{ijk}}$, i.e.,
\begin{equation}
\chi_{\Delta_{ijk}} (h_{\Delta_{ijk}} \ket{\downarrow \downarrow \downarrow}) = \chi_0 (h_{\Delta_{ijk}}\ket{\downarrow \downarrow \downarrow}), ~  \chi_0 = \frac{3\sqrt{3}}{2}.
\end{equation}
Hence, the action of magnetization-non-conserving term on the FM state of a single triangular plaquette is to generate orbital chirality. 
Since each edge on the triangular lattice is shared by exactly two plaquettes, the magnetization-non-conserving term in the Hamiltonian can be written as $\tilde{H} = \frac{J_+}{2} \sum_{\Delta} \left( h_{\Delta_{ijk}} + \mbox{h.c.} \right)$, implying that imaginary time-evolution by $\tilde{H}$ will naturally induce orbital chirality. 

To find $\langle \chi_\Delta(\tau) \rangle$ at small $\tau$, we approximate $\ket{\psi_\tau} \approx (1 - \tau H)\ket{\psi_{\rm FM}}$, which is translation invariant due to the translation invariance of $H$ and the initial state. 
Noting that $\chi_{\Delta}\ket{\psi_{\rm FM}} = 0$, the expectation value simplifies to 
$\langle \chi_\Delta(\tau)  \rangle \approx \tau^2 \bra{\psi_{\rm FM}} H \chi_{\Delta} H \ket{\psi_{\rm FM}}$.
Since $\ket{\psi_{\rm FM}}$ is an eigenstate of the magnetization-conserving terms in $H$, the only term which can induce orbital chirality is the $J_+$ term. 
Its action on $\ket{\psi_{\rm FM}}$ can be written as
\begin{equation}
J_+ \sum_{\langle ij \rangle} ( \gamma_{ij} \sigma^+_i \sigma^+_j + \mbox{h.c.})\ket{\psi_{\rm FM}} = 2^2 J_+ \sum_{\langle ij \rangle} \gamma_{ij} \ket{\uparrow_i \uparrow_j},
\end{equation}
where $\ket{\uparrow_i \uparrow_j}$ is defined to be the state with two up-spins on the NN sites $i$ and $j$ with down-spins everywhere else. 
To compute $\langle \chi_\Delta(\tau) \rangle$, the only relevant edges $\langle ij \rangle$ are the ones that belong to one of the three neighboring triangles that share an edge with $\Delta$ (including the three edges on $\Delta$ itself). 
Evaluating the expectation value on the three edges that belong to $\Delta$ (each gives $\chi_0/3$) and the six edges of the triangles adjacent to $\Delta$ (gives $\chi_0/3$ for the two edges on each adjacent triangle) leads to
\begin{equation}
\langle \chi_\Delta(\tau) \rangle = \bra{\psi_\tau} \chi_\Delta \ket{\psi_\tau} \approx  (4 J_+ \tau)^2 (2\chi_0).
\end{equation}
Thus, we have shown that the FM state after a small imaginary time evolution develops a finite orbital chirality due to the $J_+$ term. 

\begin{figure}
    \centering
    \includegraphics[scale=1.0]{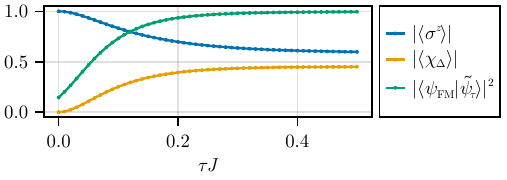}
    \caption{Imaginary time evolution of the state $\ket{\psi_{\mathrm{FM}}}$ by the Hamiltonian using ED for $N=16$. We observe that the chirality operator immediately acquires an expectation value and quickly saturates to its ground state value, whereas the magnetization is reduced. The couplings are chosen inside the chiral ferromagnetic phase, $J = 1.0, ~J_z = 0.4, ~J_+ = 0.9$. 
    }
    \label{fig:ite2}
\end{figure}

We note that the state $\ket{\psi_\tau}$ is not normalized since imaginary time evolution is not unitary. 
The normalization does not affect our argument, but is tedious analytically.
We compute this imaginary time evolution numerically with properly normalized wavefunctions $|\tilde{\psi}_\tau\rangle$ at each time step and show that the chirality saturates to its ground state value around $\tau \approx 1/2J$ [Fig.~\ref{fig:ite2}].

\subsection{Chiral stripe state}

\begin{figure}
    \centering
    \includegraphics[scale=1.0]{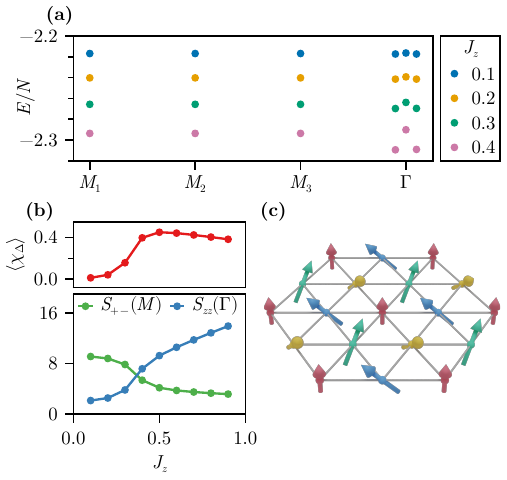}
    \caption{\textbf{(a)} The lowest energy levels of the 28-site cluster at the the three $M$ points and the $\Gamma$ point. 
    Increasing $J_z$ at fixed $J_+ = 1.25$ enhances the splitting between the lowest energy state at the $\Gamma$ point and the states at the $M$ points. 
    \textbf{(b)} [$J_+ = 1.25$] As $J_z$ is increased and the state moves from the stripe to the chiral stripe, the ED ground state acquires chirality. 
    Simultaneously, the in-plane pseudospin correlation $S_{+-}(k)$ has peaks of equal magnitude at the three $C_{3z}$-related $M$ points near the transition, while the $z$-magnetization correlation $S_{zz}(k) = \langle \sigma^z_k \sigma^z_{-k} \rangle$ is peaked at the $\Gamma$ point.
    On increasing $J_z$, the spectral weight at the $M$ points get suppressed, while uniform magnetization correlations get enhanced, indicative of a transition to a FM phase. 
    \textbf{(c)} A representative soft-pseudospin configuration for the chiral stripe phase, with uniform $z$-magnetization and in-plane ordering at all three $M$ points. 
    The red spins point in the $+\hat z$-direction, and each of the others is canted between $+\hat z$ and one of the lattice vectors.}
    \label{fig:chiralstripe}
\end{figure}

In the main text, we discussed the possible emergence of a chiral stripe phase near the phase boundary of the chiral FM and the collinear FE phases. 
Here, we present exact diagonalization (ED) data in support of our claim, discuss the associated pseudospin correlations, and argue that a classical pseudospin configuration (dense skyrmion crystal) is able to capture their key aspects.
Further, we elaborate on certain key terms in the Landau free energy functional, permitted by the $C_{3v}$ point group, that allow such a chiral stripe to emerge as a low-energy phase. 

The low-energy ED spectrum as a function of $J_z$ at fixed $J_+$, traversing the stripe-FM phase boundary obtained from VUMPS [Fig.~\ref{fig:chiral_fm}(a)], is shown in Fig.~\ref{fig:chiralstripe}(a) for $N = 28$ sites~\cite{fnED}. 
While the three collinear stripe phases, corresponding to the three $M$-point states in Fig.~\ref{fig:chiralstripe}(a), are degenerate as expected, they are higher in energy than the two ground states at the $\Gamma$ point. 
Further, the two $\Gamma$-point states are related by time-reversal: one is chiral [Fig.~\ref{fig:chiralstripe}(b)] and the other is anti-chiral. 
For both these states, the in-plane correlations $\langle \sigma^+_0 \sigma^-_i\rangle$ exhibit equally strong peaks at all three $M$ points [Fig.~\ref{fig:chiralstripe}(b)], in contrast with VUMPS where $M$ point correlations are not $C_{3z}$-symmetric even in the FM [Fig.~\ref{fig3}(b)]. 
However, the out-of-plane $\langle \sigma^z_0 \sigma^z_i\rangle$ correlations remain peaked at the $\Gamma$ point. 
These observations hint that an intermediate chiral phase which breaks translation symmetry could exist in the thermodynamic limit, and ED simply restores translation symmetry by averaging over all sectors to bring the state to the $\Gamma$ point. 
Since this phase features both orbital chirality and AFE stripe-like correlations, we call it the chiral stripe. 

The pseudospin configuration corresponding to the chiral stripe [Fig.~\ref{fig:chiralstripe}(c)] may be obtained by analyzing the Landau free energy density, as mentioned in the main text. 
While a detailed analysis including all possible symmetry-allowed terms is presented in the SM~\cite{SM}, here we focus on the key terms for the $\psi_n$ fields [Eq.~\eqref{eq:softfields}] that describe in-plane ordering at $M$ points. 
\begin{align}
 F^{(2)} \ni & \; r \left( \sum_{n=1}^{3} |\psi_n|^2 \right) + \lambda \, \text{Re}\big[\psi_1^2 + \omega \psi_2^2 + \omega^2 \psi_3^2\big]  \nonumber \\
F^{(4)} \ni  & \;  u_1 \left( \sum_n |\psi_n|^4 \right) + u_2 \left(\sum_{n < n'} |\psi_n|^2 |\psi_{n'}^2| \right) 
\end{align}
Using the amplitude-phase representation for the complex field $\psi_n = |\psi_n| e^{i \phi_n}$, we note that the $C_{3v}$ symmetry allows an additional anisotropic term in the quadratic free energy $F^{(2)}$ proportional to $\lambda$.
This term, typically disallowed by an internal $O(2)$ symmetry in spin Hamiltonians~\cite{Jin2025}, is enabled here by the orbital nature of the pseudospins and explicitly prefers locking of the phases: $\phi_n  = 2 (n-1) \pi/3$ at the free energy minimum (assuming $\lambda < 0$). 
Further, for $u_2 > 2 u_1$ in the quartic term $F^{(4)}$, the mutual repulsion between the different $\psi_n$ fields is stronger than their self repulsion, and the minimum for the free energy is obtained when $|\psi_1| = |\psi_2| = |\psi_3|$. 
Lastly, while our discussion thus far has not accounted for the magnetization soft-field $m_0$---an attractive quartic coupling between $m_0$ and $\psi_n$ is symmetry-allowed. 
This coupling prefers coexistence of both orders, while retaining the phase-locking for the $\psi_n$-fields at the free energy minimum (see SM~\cite{SM} for more details).
Taken together, we find a (soft) pseudospin configuration given by
\begin{align}
\langle \bm{\sigma}_i \rangle = &   \sum_{n=1}^{3} \cos[ \M_n \cdot \r_i + \phi_n] \, \hat{x} +  \sum_{n=1}^{3} \sin[ \M_n \cdot \r_i + \phi_n] \, \hat{y} \nonumber \\ 
& + m_0 \, \hat{z}, 
\end{align}
as schematically depicted in Fig.~\ref{fig:chiralstripe}(c).
Such a pseudospin configuration breaks translation symmetry with an enlarged four-site unit cell, but otherwise features analogous in-plane and out-of-plane pseudospin correlations as the ED ground state in the vicinity of the stripe-FM phase boundary.

\onecolumngrid
\newpage

\def \r{{\bm r}}
\def \q{{\bm q}}
\def \k{{\bm k}}
\setcounter{secnumdepth}{3}
\numberwithin{equation}{section}
\renewcommand\theequation{S\arabic{section}.\arabic{equation}}
\renewcommand{\thefigure}{S\arabic{figure}}
\renewcommand{\thetable}{S\arabic{table}}
\setcounter{figure}{0}
\setcounter{table}{0}
\renewcommand{\bibnumfmt}[1]{[S#1]}
\renewcommand{\citenumfont}[1]{S#1}




\begin{center}
    \textbf{Supplemental Material: Fluctuation-driven chiral ferromagnetism}
\end{center}


\section{Lattice conventions and model symmetries}

\begin{figure}[!h]
    \centering
    \includegraphics[width=0.5\linewidth]{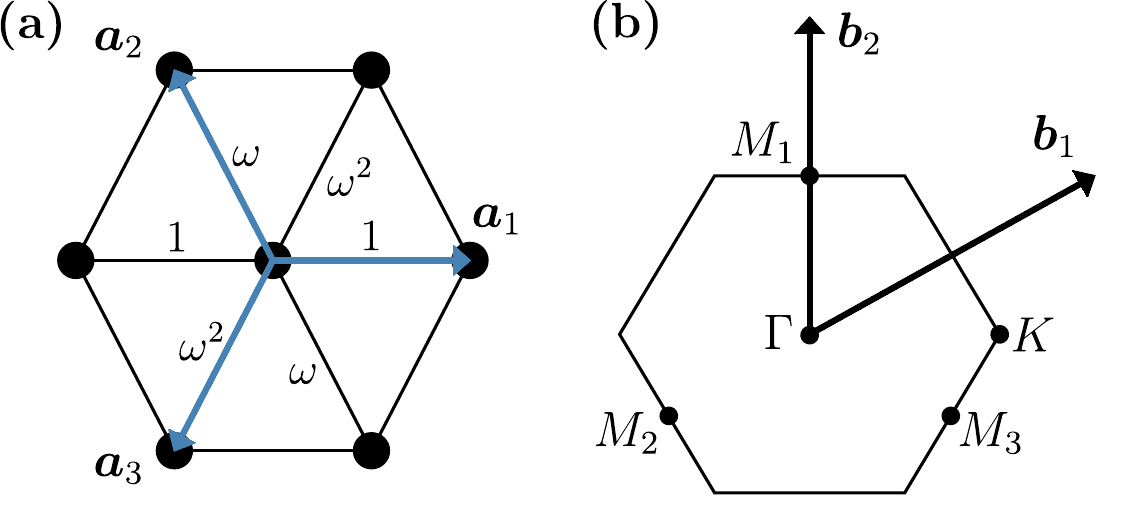}
    \caption{(a) The triangular lattice along with the nearest neighbor vectors $\pm \bm{a}_{1,2,3}$. The phases $\gamma_{ij} = 1, \omega, \omega^2$ for the magnetization non-conserving $\sigma^+_i \sigma^+_j$ term in the Hamiltonian are shown on the corresponding bonds, where $\omega = e^{2\pi i /3}$. 
    (b) The BZ along with reciprocal lattice vectors. The high symmetry points are also shown.
    The coordinates of the three $M$ points are set by $\M_1 = \b_2/2$, $\M_2 = - \b_1/2$ and $\M_3 = (\b_1 - \b_2)/2$.
    }
    \label{fig:lat}
\end{figure}

In this section, we elaborate on the setup, the symmetries of our Hamiltonian $H$ [Eq.~\eqref{eq:H} in the main text], and the corresponding symmetry transformations of the pseudospin operators $\bm{\sigma}_i$. 
We work with the triangular lattice [Fig.~\ref{fig:lat}(a)] where we take the two basis vectors as $\bm{a}_1 = a(1, ~0)$ and $\bm{a}_2 = a(-1/2, ~\sqrt{3}/2)$, and define $\a_3 = - (\a_1 + \a_2)$. 
The basis vectors of the hexagonal reciprocal lattice [Fig.~\ref{fig:lat}(b)] are $\bm{b}_1 = \frac{2\pi}{a} (1, ~1/\sqrt{3})$ and $\bm{b}_2 = \frac{2\pi}{a} (0, ~2/\sqrt{3})$.  
For convenience, we set $a = 1$, unless otherwise mentioned.

Our triangular lattice Hamiltonian $H$ is symmetric under translations $T_{\a_1}$ and $T_{\a_2}$, and possesses the point group symmetry of $C_{3v}$, which is generated by $2\pi/3$ rotations $C_{3z}$ and the mirror $M_y$. 
In addition, it is symmetric under time-reversal $\mathcal{T}$, and further exhibits an internal $\mathbb{Z}_2$ symmetry corresponding to flipping the in-plane components of $\bm{\sigma}_i$ via $U(\hat{z},\pi) = \prod_{i} \sigma^z_i$, corresponding to a global $\pi$ pseudospin rotation about $\hat{z}$. 
These symmetry transformations are summarized in Table~\ref{tab:sym}. 
\begin{table}[h!]
\begin{center}
\begin{tabular}{ |c|| c| c|} 
 \hline
 \multirow{2}{*}{Symmetry}&  \multicolumn{2}{c|}{Pseudospin}\\
& ${\sigma}^+_i$ &  $\sigma^z_i$
\\
\hline \hline 
 \rule{0pt}{3ex}$C_{3z}$ &  $\omega^2 \sigma^+_{C_{3z} \bm{r}_i}$ & $\sigma^z_{C_{3z} \bm{r}_i}$ \\ 
 \rule{0pt}{4ex}$M_y$ &  $ \sigma^-_{M_y \bm{r}_i}$ & $- \sigma^z_{M_y \bm{r}_i}$  \\ 
 \rule{0pt}{4ex}$\mathcal{T}$  & $\sigma^{-}_i$ &  $- \sigma^z_i$ \\
 \rule{0pt}{4ex}$\mathbb{Z}_2$  & $-\sigma^+_i$ & $\sigma^z_i$
 \\
 \hline 
\end{tabular}
\caption{Point group and internal symmetry actions on the orbital pseudospin operators $\sigma^+_i = \sigma^x_i + i \sigma^y_i$ and $\sigma^z$. 
$C_{3z}$ rotates the vector $\bm{r}_i$ by $2\pi/3$ about the $z$-axis and also adds an internal phase $\omega^2$ to $\sigma_i^+$.
$M_y \bm{r_i} = M_y(x_i, y_i) \equiv (-x_i,y_i)$ is a spatial reflection. 
The $\mathbb{Z}_2$ spin-flip symmetry is generated by the operator $U(\hat{z},\pi) = \prod_{i} \sigma^z_i$, which satisfies $U^2 = \mathbb{I}$. $\mathcal{T}$ represents time reversal symmetry.}
\label{tab:sym}
\end{center}
\end{table}

\section{Classical Energy (Luttinger-Tisza analysis)}

In this section, we provide the details of the Luttinger-Tisza (LT) analysis that is used in the main text.
The method represents quantum pseudospins as classical two-dimensional vectors of unit length, with a Hamiltonian featuring anisotropic pseudospin interactions that take the general form,
\begin{equation}
    H = \frac{1}{2}\sum_{i,j} J_{i j}^{\alpha \beta}\sigma_{i}^\alpha \sigma_j^\beta,
\end{equation}
on any Bravais lattice. 
We proceed by writing the Hamiltonian in momentum space by taking
\begin{equation}
    \bm{\sigma}_{i} = \frac{1}{\sqrt{N}}\sum_{\bm{k}}\tilde{\bm{\sigma}}_{\bm{k}} \, e^{i\bm{k}\cdot\bm{R}_i}.
\end{equation}
The Hamiltonian can be written as  
\begin{equation}
    \label{eq:LTHam}
    H = \sum_{\bm{k}} \mathcal{J}_{\alpha \beta}(\bm{k})\tilde{\bm{\sigma}}_{-\bm{k}}^\alpha \tilde{\bm{\sigma}}_{\bm{k}}^{\beta},
\end{equation}
where $\mathcal{J}^{\alpha \beta}(\bm{k})$ is the Fourier-transformed interaction matrix. 
The problem is to find a spin configuration that minimizes this quadratic form subject to the strong constraint that all spins be unit-norm: this constraint usually makes the problem analytically intractable. 
The Luttinger-Tisza method relaxes this constraint to the weak constraint
\begin{equation}
    \sum_{i} |\bm{\sigma}_{i}|^2 = N,
\end{equation}
or equivalently in momentum space,
\begin{equation}
    \sum_{\bm{k}} |\bm{\sigma}_{\bm{k}}|^2 = N,
\end{equation}
where $N$ is the number of unit cells.  
This weak constraint is a necessary (but not sufficient) condition for the the strong constraint to be satisfied.
With it, though, the problem of minimization simplifies into a momentum-diagonal eigenvalue problem for the quadratic form in Eq.~\ref{eq:LTHam}. If the solution to that problem also happens to satisfy the strong constraint, then we know that it is a valid classical ground state.

It is useful to write $H$ (Eq.~\eqref{eq:H} in the main text) in Cartesian form \cite{Iaconis2018spin}
\begin{equation}
   H = -\sum_{\langle ij \rangle} \left[J_{z}\sigma_i^z \sigma_j^z + J_{xx}(\sigma_i^x \sigma_j^x + \sigma_i^y \sigma_j^y)  + 4J_{+}(\bm{\sigma}_i \cdot \hat{e}_{ij})(\bm{\sigma}_j \cdot \hat{e}_{ij})\right],
\end{equation}
where $J_{xx} = J- 2J_{+}$ and $\hat{e}_{ij}$ is the unit vector pointing from site $i$ to site $j$. 

Then the Fourier transform of the interaction matrix of our Hamiltonian takes the following 3 by 3 matrix form

\begin{equation}
    \label{eq:J(q)}
    \mathcal{J}(\k) = \sum_{\mu=1}^{3} [\mathbf{J}(\hat{e}_\mu)\cos(\k\cdot \bm{a}_\mu)], \; \text{ where }
    \mathbf{J}(\hat{e}) = -\begin{pmatrix}
        J_{xx}+4J_{+}e_x^2 & 4J_{+}e_x e_y & 0 \\
        4J_{+}e_x e_y & J_{xx}+4J_{+}e_y^2 & 0 \\
        0 & 0 & J_{z}
    \end{pmatrix}
\end{equation}

We find that the minimum of $\mathcal{J}(\k)$ occurs either at the $\Gamma$ point, or at the $M$ points as the parameters $J_+$ and $J_z$ are varied, keeping $J = 1$ fixed. 
On demanding that the strong LT constraint is satisfied, we get either a ferroelectric (FE) or a ferromagnet (FM) in the former case, and an antiferroelectric stripe in the latter case. 
We analyze these scenarios in turn. 

\emph{FE} and \emph{FM}: We set $\k = 0$ in Eq.~\eqref{eq:J(q)} to get 
\begin{equation}
    \mathcal{J}(0) = -\begin{pmatrix}
        3J_{xx}+6J_{+} & 0 & 0 \\
        0 & 3J_{xx}+6J_{+} & 0 \\
        0 & 0 & 3J_{z}
    \end{pmatrix} = -\begin{pmatrix}
        3J & 0 & 0 \\
        0 & 3J & 0 \\
        0 & 0 & 3J_{z}
    \end{pmatrix}
\end{equation}
Note that we used the fact that $J_{xx} = J - 2J_{+}$. We see that this matrix has two degenerate minimum eigenvalues at $\lambda = -3J$ when $J > J_{z}$. There are two degenerate eigenvectors that can point either in the $x$ or the $y$ direction thus the state can point anywhere in the plane with the same energy. The classical state has accidental U(1) symmetry which we will see later gets broken by quantum fluctuations which can be captured at the level of linear spin wave theory (LSWT). Note that when $J_{z} > J$, the minimum is unique and corresponds to $\lambda = - 3 J_z$, which is the z-polarized FM state.  \\

\emph{Stripe}: Now we choose one of the M points, $\k_M = \left(0, \frac{2\pi}{\sqrt{3}}\right)$. 

\begin{equation}
    \mathcal{J}(\k_M) = \begin{pmatrix}
        J_{xx}-2J_{+} & 0 & 0 \\
        0 & 6J_{+}+J_{xx} & 0 \\
        0 & 0 & J_{z}
    \end{pmatrix} = \begin{pmatrix}
        J-4J_{+} & 0 & 0 \\
        0 & J+4J_{+} & 0 \\
        0 & 0 & J_{z}
    \end{pmatrix}
\end{equation}

The minimum eigenvalue corresponding to the stripe is found to be $\lambda = J-4J_{+}$, with the corresponding eigenvector pointing in the $\hat{x}$ direction. 
A general solution is $\bm{\sigma}(\bm{r}_i) = \text{Re}[\bm{v} \, e^{i\k\cdot\bm{r}_i}]$, which in this case reduces to $\bm{\sigma}(\bm{r}_i) = \cos(\k_M \cdot\bm{r}_i)\hat{x}$. 
For the $M$ point, the cosine factor is either $1$ or $-1$ at any point on the lattice thus this is a constant length spin vector (strong constraint is satisfied) that is alternating in a stripy fashion. 
In contrast with the FE, there is no accidental U(1) degeneracy as the stripes are locked to the lattice. 
Furthermore, we can determine that the transition between the FE and the stripe happens when the two states are degenerate, i.e., $J - 4 J_+ = - 3 J$, or equivalently, when $J_{+} = J$.

Now at $J_{+} = J$, consider $\q = \lambda \k_M$ where $0 < \lambda < 1$ which is a point along the line connecting $\Gamma$ to $\k_M$. This gives 
\begin{equation}
    \mathcal{J}(\q) = -\begin{pmatrix}
        3J & 0 & 0 \\
        0 & J(4\cos\lambda\pi-1) & 0 \\
        0 & 0 & J_{z}(2\cos\lambda\pi+1)
    \end{pmatrix}
\end{equation}

We see that at $J_{+} = J$, all momenta along the lines connecting $\Gamma$ to the M points have the same LT optimal eigenvalue $-3J$ which is the weak LT degeneracy discussed in the main text. 

\paragraph{Real-space optimization.} The Luttinger-Tisza procedure is only guaranteed to find the classical ground state in simple cases. To verify its results by other means, we also perform numerical optimization of the classical energy. We start from the assumption that the ground state configuration retains some subgroup of the $\mathbb Z \times \mathbb Z$ translational symmetry group of the triangular lattice, reducing the size of the manifold of valid configurations to $ (S^2)^n$, where $n$ is the number of sublattices. We then use manifold conjugate gradient descent to find the lowest-energy configurations \cite{kofod2025,edelman1998,absil2008, KVMC2026}.  This is not a convex optimization problem, so we do not have robust guarantees for finding global minima, but when there are not too many sublattices, we expect to find the true minimum by running the optimization many times from random initial states. Looking at symmetry-breaking patterns with up to 12 sublattices, we find agreement with the Luttinger-Tisza results in the whole diagram, corroborating this analysis.

\section{Spin Wave Excitations}
\label{SMsec:SWT}
In this section, we provide a detailed derivation of the low-energy excitations of the different classically ordered phases. 
For the ferroelectric, we will use our results to compute the zero point energy due to quantum fluctuations, and argue that this will reduce the accidental $U(1)$ symmetry of the ground state and lock the pseudospins to the lattice vectors. 
For the ferromagnet, we will use our results to calculate the thermal Hall conductivity at finite temperatures.  

\subsection{Ferroelectric}
We use the vacuum predicted classically where the spins order in the $xy$ plane. 
We consider the general case of spins pointing at an angle $\phi$ in the xy plane, since these are all degenerate in our LT analysis.
To carry out the linear-spin wave analysis, it is convenient to define a locally rotated coordinate system $(x', y', z')$, such that the classical ground state spin configuration is $\langle \bm{\sigma}_i \rangle = - \hat{\mathbf{z}}'$ for all sites. 
In terms of the original lab-fixed frame, the orthogonal basis vectors in the rotated frame are given by:
\begin{align}
    \hat{\mathbf{z}}' &= (\cos\phi, \sin\phi, 0), \nonumber \\
    \hat{\mathbf{x}}' &= (\sin^2\phi, -\sin\phi\cos\phi, -\cos\phi), \nonumber \\
    \hat{\mathbf{y}}' &= (-\sin\phi\cos\phi, \cos^2\phi, -\sin\phi).
\end{align}
The spin operators in the original laboratory frame $(x,y,z)$ can be expressed in terms of the local frame operators $(\sigma^{x'}, \sigma^{y'}, \sigma^{z'})$ as:
\begin{align}
    \sigma_i^z &=  -\cos\phi ~\sigma_i^{x'} - \sin\phi ~\sigma_i^{y'}, \nonumber \\
    \sigma_i^x &= \sin^2\phi ~\sigma_i^{x'} - \sin\phi \cos\phi ~\sigma_i^{y'} + \cos\phi ~\sigma_i^{z'}, \nonumber \\
    \sigma_i^y &= -\sin\phi \cos\phi ~\sigma_i^{x'} + \cos^2\phi ~\sigma_i^{y'} + \sin\phi ~\sigma_i^{z'}.
\end{align}
Next, we introduce bosonic creation and annihilation operators $a_i^\dagger, a_i$ via the Holstein-Primakoff transformation~\cite{Auerbach_Book}:
\begin{align}
    \sigma_i^{z'} &= 2a_i^\dagger a_i - 1, \nonumber \\
    \sigma_i^{x'} &= (a_i^\dagger + a_i), \nonumber \\
    \sigma_i^{y'} &= i (a_i - a_i^\dagger).
\end{align}
Substituting these into the spin operators, we find
\begin{align}
   \sigma_i^z & =  -(e^{-i\phi}a_i^\dagger + e^{i\phi}a_i), \nonumber \\
    \sigma_i^+ &= e^{i\phi}\left[2a_i^\dagger a_i - 1 + \left( e^{-i\phi}a_i^\dagger - e^{i\phi}a_i \right) \right]\nonumber \\
    \sigma_i^- &= e^{-i\phi}\left[2a_i^\dagger a_i - 1 - \left( e^{-i\phi}a_i^\dagger - e^{i\phi}a_i \right)\right] .
\end{align}

\begin{figure}
    \centering
    \includegraphics[width=1.0\linewidth]{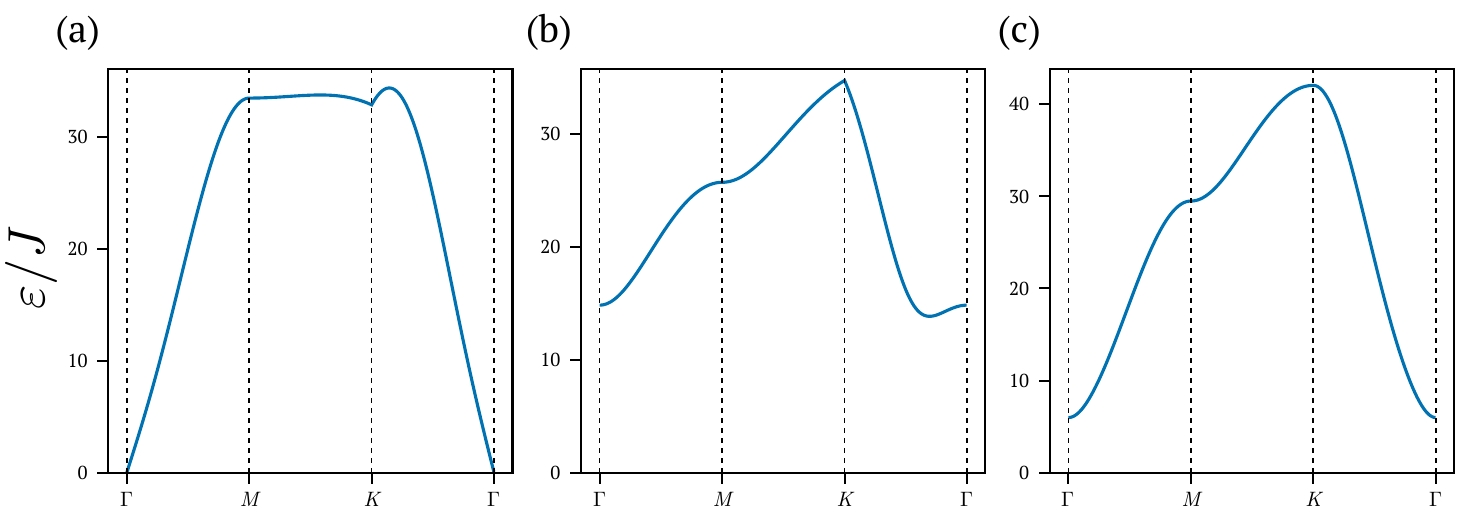}
    \caption{Representative plots for the magnon spectrum of the three classical states. (a) The FE magnon band ($J_z = 0.5, ~J = 1.0, ~ J_+ = 0.5$). It is gapless at the gamma point. (b) The stripe magnon band for ($J_z = 0.5, ~J = 1.0, ~ J_+ = 1.5$). (c) The FM magnon band for ($J_z = 1.5, ~J = 1.0, ~ J_+ = 0.5$). Both the stripe and the FM bands are gapped.}
    \label{fig:SWspec}
\end{figure}


We now express each of the terms in our Hamiltonian to quadratic order in the Holstein-Primakoff boson operators. 
\begin{equation}
    -J_z \sum_{\langle ij \rangle} \sigma_i^z \sigma_j^z = -J_z \sum_{\langle ij \rangle} \left[e^{2i\phi}a_i a_j + a_i a_j^\dagger + a_i^\dagger a_j + e^{-2i\phi}a_i^\dagger a_j^\dagger \right].
\end{equation}
Next, we consider terms of the form $\sigma_i^+ \sigma_j^-$. 
\begin{align}
    \sigma_i^+ \sigma_j^- &= \left[ 2a_i^\dagger a_i - 1 +  \left( e^{-i\phi}a_i^\dagger - e^{i\phi}a_i \right) \right] 
    \left[ 2a_j^\dagger a_j - 1 -  \left( e^{-i\phi}a_j^\dagger - e^{i\phi}a_j \right) \right] \nonumber \\
    &= -2(a_i^\dagger a_i + a_j^\dagger a_j) - \left( e^{-i\phi}a_i^\dagger - e^{i\phi}a_i \right)\left( e^{-i\phi}a_j^\dagger - e^{i\phi}a_j \right) + \mathcal{O}(a^4)
\end{align}
Note that linear and cubic terms sum to zero. 
Finally, we consider terms of the form $\sigma_i^+ \sigma_j^+$:
\begin{align}
    \sigma_i^+ \sigma_j^+ &= e^{2i\phi}\left[ 2a_i^\dagger a_i - 1 + \left( e^{-i\phi}a_i^\dagger - e^{i\phi}a_i \right) \right] 
    \left[ 2a_j^\dagger a_j - 1 + \left( e^{-i\phi}a_j^\dagger - e^{i\phi}a_j \right) \right] \nonumber \\
    &= e^{2i\phi}\left( e^{-i\phi}a_i^\dagger - e^{i\phi}a_i \right)\left( e^{-i\phi}a_j^\dagger - e^{i\phi}a_j \right) + \mathcal{O}(a^3)
\end{align}
Note that other chemical potential, and linear terms sum up to zero when including the bond-dependent phases. 
We also make a unitary transformation $a_i \rightarrow e^{-i\phi} a_i$ which cancels the phase in the spin conserving terms as it should but keeps the phase factor outside of $J_+$ term.   
The quadratic part of the Hamiltonian then takes the following form:
\begin{equation}
    H^{(2)}_{\mathrm{SW}} = 3J \sum_i a_i^\dagger a_i + \sum_{\langle ij \rangle} \left[\Delta_{ij} a_i a_j + t_{ij} a_j^\dagger a_i + t_{ij}^\ast a_i^\dagger a_j + \Delta_{ij}^\ast a_i^\dagger a_j^\dagger \right],
\end{equation}
where $t_{ij} = -J_{z} - 2J + \tilde{\gamma}_{ij} J_{+}$ and $\Delta_{ij} = -J_{z} + 2J - \tilde{\gamma}_{ij} J_{+}$. Here $\tilde{\gamma}_{ij} = 2\cos(2\phi+2n\pi/3)$ with $n=0,~1,~2$ for bonds along $\bm{a}_1, ~\bm{a}_2, ~\bm{a}_3$. By going into momentum space we get the following, 
\begin{equation}
    H = \frac{1}{2}\sum_{\bm{k}} \Psi_{\bm{k}}^\dagger h(\bm{k}) \Psi_{\bm{k}}, 
\end{equation}
where $\Psi_{\bm{k}} = (a_{\bm{k}} ~a^\dagger_{-\bm{k}})^T$, and $h(\bm{k})$ is given by: 
\begin{equation}
    h(\bm{k}) = \begin{pmatrix}
        A(\bm{k}) & B(\bm{k})\\
        B(\bm{k}) & A(\bm{k})
    \end{pmatrix},
\end{equation}
where $A(\bm{k}) = 3J + 2(t_1 \cos(\bm{k}\cdot\bm{a}_1) + t_2 \cos(\bm{k}\cdot\bm{a}_2) + t_3 \cos(\bm{k}\cdot\bm{a}_3))$ and $B(\bm{k}) = 2(\Delta_1 \cos(\bm{k}\cdot\bm{a}_1) + \Delta_2 \cos(\bm{k}\cdot\bm{a}_2) + \Delta_3 \cos(\bm{k}\cdot\bm{a}_3))$. 


\subsection{Stripe}

We make the choice of a stripe order $\langle \bm{\sigma} \rangle = \cos(\bm{k}_M \cdot \bm{r}) \hat{x}$ with $\bm{k}_M = (0, \frac{2\pi}{\sqrt{3}})$. Similar to the FE we need to rotate our spin operators, but here we need to do a site dependent rotation.
The calculations follow in a way similar to the FE, and we simply provide the final form of the Hamiltonian: 
\begin{equation}
    H = \frac{1}{2} \sum_{\k} \Psi_{\k}^\dagger h(\k) \Psi_{\k},
\end{equation}
where $\Psi_{\k} = (a_{\k} ~a_{-\k}^\dagger)^T$, and $h(\k)$ is 
\begin{equation}
    h(\k) = \begin{pmatrix}
        A(\k) & B(\k)\\
        B(\k) & A(\k)
    \end{pmatrix},
\end{equation}

where $A(\k) = -16J_{+}-4J-(4J_{+}+2J+2J_{z})\cos(\k\cdot \bm{a}_1) + (2J_{+}-2J+2J_{z})(\cos(\k\cdot \bm{a}_2)+\cos(\k\cdot \bm{a}_3))$ and $B(\k) = (-4J_{+}-4J+2J_{z})\cos(\k\cdot \bm{a}_1) + (2J_{+}-4J-2J_{z})(\cos(\k\cdot \bm{a}_2)+\cos(\k \cdot \bm{a}_3))$.

\subsection{Ferromagnet}

For the ferromagnet, we do not need to rotate the spin operators. Thus, we get the following Hamiltonian  
\begin{equation}
    H = \frac{1}{2} \sum _{\k} \Psi_{\bm{k}}^\dagger h(\bm{k}) \Psi_{\bm{k}}, 
\end{equation}
where $\Psi_{\bm{k}} = (a_{\bm{k}} ~a_{-\bm{k}}^\dagger)^T$ and 
\begin{equation}
    h(\bm{k}) = \begin{pmatrix}
        A(\bm{k}) & B(\bm{k})\\
        B(\bm{k})^\ast & A(\bm{k})
    \end{pmatrix},
\end{equation}
where $A(\bm{k}) = -4J (\cos(\bm{k}\cdot\bm{a}_1) + \cos(\bm{k}\cdot\bm{a}_2) + \cos(\bm{k}\cdot\bm{a}_3)) + 12J_{z}$ and $B(\bm{k}) = -8J_{+} (\cos(\bm{k}\cdot\bm{a}_1) + \omega\cos(\bm{k}\cdot\bm{a}_2) + \omega^2\cos(\bm{k}\cdot\bm{a}_3))$. Representative line-cuts for the magnon dispersion in the three classical ground states are presented in Fig. \ref{fig:SWspec}.

\section{Order by disorder}

\begin{figure}
    \centering
    \includegraphics[width=0.7\linewidth]{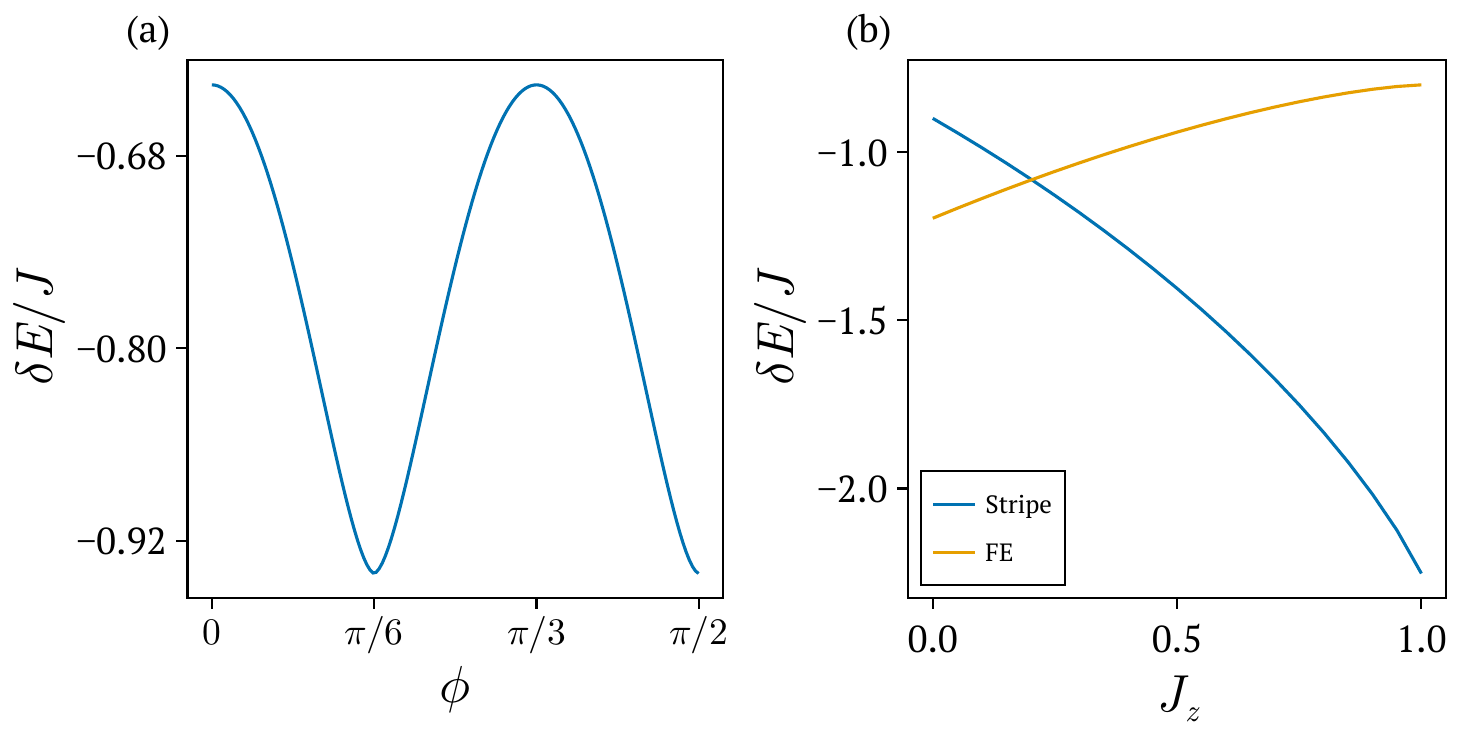}
    \caption{Order by disorder mechanisms. (a) The accidental $U(1)$ symmetry of the FE is broken in linear order in spin wave theory. States that point along the lines $\pi/6, ~\pi/2$ measured from the $x$ axis have lower energy than the rest. (b) Along the line of the classical degeneracy between the FE and the stripe ($J = J_+$ and $J_z < 1.0$) for small $J_z$ the FE has lower energy, however, above a critical $J_z$ there is a transition and the stripe has the lower energy. }
    \label{fig:FEvsS}
\end{figure}

In the ferroelectric there is an accidental classical $U(1)$ degeneracy, where the moment $\langle \sigma_i \rangle$ can point anywhere in the xy plane.
This is also the reason why the magnon dispersion in this state is gapless. However, the symmetry of the Hamiltonian $H$ is reduced to $C_{3v}$ by the $J_+$ term so we do not expect this accidental classical degeneracy to survive quantum fluctuations. 
Here, using the magnon dispersion derived in the previous section, we provide a direct computation of the quantum zero-point energy in linear spin-wave theory:
\begin{equation}
    \delta E = \frac{1}{2} \sum_{\k} [\varepsilon(\k) - A(\k)]
\end{equation}
The results of our computation are shown in Fig. \ref{fig:FEvsS}. 
We find that the ordered states where $\langle \bm{\sigma} \rangle$ points perpendicular to the lattice principal directions have lower energy, implying the $U(1)$ accidental symmetry is reduced to $Z_6$ as stated in the main text.

Another form of order by disorder occurs along the classical phase boundary ($J_+ = J$) between the FE and the stripe. 
We find that for $J_z = 0$ the energy of the FE is smaller than the stripe, and this continues to be the case for small $J_z$ along the line $J_+ = J$.
However, beyond a critical $J_z \approx 0.2$ the stripe is energetically favored till the point where the three phases meet classically. 
Interestingly, at the classical tri-critical point $J_+ = J_z = J$, the FM wins the competition.
This may be intuitively understood as an order by disorder mechanism by considering the magnon dispersion for each of these three states. 
While both the FE and the stripe, the magnon is gapless only at the $\Gamma$ point, the magnon dispersion for the FM vanishes along the lines connecting the $\Gamma$ point to the $M$ points in the BZ --- leading to lower zero-point energy for the FM.
Interestingly, these are the same degeneracy lines that we encountered in our LT analysis, corresponding to the minima of $\mathcal{J}(\k)$ along $J_+ = J, ~ J_z < 1$: however, this vanishing is restricted to the tri-critical point once quantum fluctuations are incorporated via linear spin-wave theory.  

\section{Magnon Berry Curvature}
In this section, we provide the details on the computation of the magnon-Berry curvature in the ferromagnet. 
As derived previously in section~\ref{SMsec:SWT} C, the $2 \times 2$ Hamiltonian $h(\k)$ in Nambu space is given by
\begin{equation}
    h(\bm{k}) = \begin{pmatrix}
        A(\bm{k}) & B(\bm{k})\\
        B(\bm{k})^\ast & A(\bm{k})
    \end{pmatrix},
\end{equation}
where $A(\bm{k}) = -4J (\cos(\bm{k}\cdot\bm{a}_1) + \cos(\bm{k}\cdot\bm{a}_2) + \cos(\bm{k}\cdot\bm{a}_3)) + 12J_{z}$ and $B(\bm{k}) = -8J_{+} (\cos(\bm{k}\cdot\bm{a}_1) + \omega\cos(\bm{k}\cdot\bm{a}_2) + \omega^2\cos(\bm{k}\cdot\bm{a}_3))$.

The Berry connection in the $n$th magnon band is given by
\begin{equation}
    \mathcal{A}^{(n)}_\mu(\k) = i[\eta U^\dagger(\k) \eta \partial_{\mu} U(\k)]_{nn},
\end{equation}
where $\eta = \mathrm{diag}(1,~-1)$ and $U(\k)$ is the paraunitary matrix that diagonalizes $h(\k)$. The Berry curvature is in turn given by 
\begin{equation}
\Omega^{(n)}(\k) = \partial_{k_x}\mathcal{A}^{(n)}_y(\k) - \partial_{k_y}\mathcal{A}^{(n)}_x(\k).  
\end{equation}
The results of this computation are presented in Fig.~\ref{fig:FMTHE} in the main text.

\section{Tensor network computational details}
All of our tensor network calculations are done using the VUMPS algorithm from the \texttt{MPSKit.jl} library \cite{zauner-stauber_variational_2018, vandammeMPSKit2025}.  We study infinite YC-wrapped cylinders with circumferences $L_y = 4, 6$, and $8$ and bond dimensions $D = 1024, 1536, 2048, 2304$, and measure convergence by means of the Galerkin error, the norm of the component of the energy gradient normal to the manifold of matrix product state of fixed $D$ \cite{vanderstraeten2019}.  We typically see convergence around $\epsilon \lesssim 10^{-5}$ for $L_y = 4$ and $\epsilon \lesssim 10^{-3}$ for $L_y = 6$. In this section, we present the results that demonstrate the claims in the main text.

\subsection{First-order FE/FM transition}
\begin{figure}
    \centering
    \includegraphics[scale=1.0]{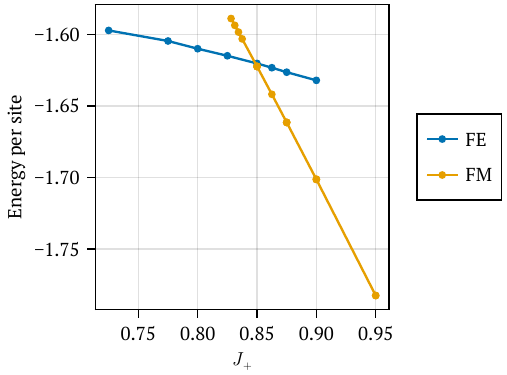}
    \caption{Along $J_z = 0.4$, there is a first-order phase transition between the ferroelectric (FE) and  ferromagnetic (FM) phases at $J_+ = 0.85$. VUMPS is able to converge on metastable states around this point, so the true ground state energy is not differentiable at the transition. FE orders at angle $\phi =\pi/6$, for comparison with Fig.~\ref{fig:FEvsS}, which is consistent with the order-by-disorder picture from spin wave theory. Data taken at $L_y = 6, D = 1536 - 2304$.}
    \label{fig:fe_trans}
\end{figure}

Except for at the ferromagnetic Heisenberg point $J_z = 1, J_+ = 0 $, the transition between the in-plane ferroelectric and the ferromagnet is first-order. This manifests in VUMPS calculations as a point of non-differentiability in the ground state energy.  Additionally, by initializing a VUMPS run in the ground state of the ``wrong'' phase, we see that it is able to converge on metastable states with higher energy than the ground state [Fig.~\ref{fig:fe_trans}].  This is in line with the prediction from Landau theory, as the FM phase breaks $\mathcal T$ and the FE breaks $C_{3z}$: as such, a second-order phase transition between these phases is Landau forbidden without fine-tuning.

\subsection{Phase diagram with small DMI}
A Dzyaloshinskii-Moriya (DM) interaction appears in the Hamiltonian of Eq.~\eqref{eq:H} as an imaginary part of $J$.  Applying a small DM term has no effect on the classical Luttinger-Tisza analysis until $\Im[J] > \sqrt{3} \Re[J]$, at which point the FE phase is replaced by a $K$-point antiferroelectric --- the orbital equivalent of the $120^\circ$-ordered N\'eel antiferromagnet for spins. 

\begin{figure}
    \centering
    \includegraphics[scale=1.0]{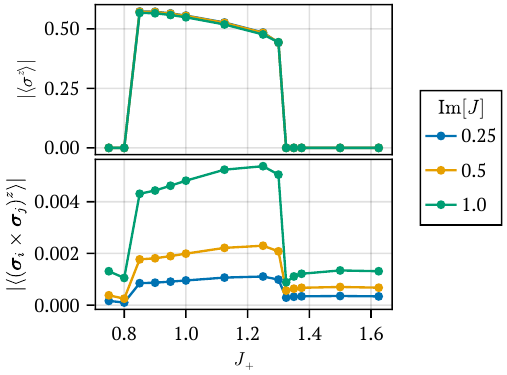}
    \caption{At $J_z = 0.4$, even a DM coupling as large as $\Im[J] = 1$ results in a small vector chirality $\Im \ev{\sigma_i^+ \sigma_j^-} = \ev{(\bm{\sigma}_i \times \bm{\sigma}_j)^z}$ on nearest neighbors and no significant change in magnetization, chirality, energy, or magnitude of the stripe or FE order parameters. Data taken at $L_y=4, D = 1536$. }
    \label{fig:dmi}
\end{figure}

In the quantum model, the phase diagram from VUMPS remains the same on the $J_z = 0.4$ cut [Fig.~\ref{fig:dmi}] and in the tip of the ferromagnetic wedge at $(J_z, J_+) = (0.25, 1)$ under the addition of nonzero $\Im[J]$---the changes to the order parameters are very slight, and the vector chirality $(\bm{\sigma}_i \times \bm{\sigma}_j)^z$ generated as a linear response to the DM term is small and still unable to account for the three-site scalar chirality as a decoupling into $\langle \chi_{ijk} \rangle \sim \langle \sigma^z_i \rangle \langle (\bm{\sigma}_j \times \bm{\sigma}_k)^z \rangle$.  
This implies that the chiral physics of this ferromagnet is not specific to the $C_{6v}$-symmetric case where $\Im[J] = 0$.
Further, it also does not rely on the DM mechanism that gives rise to scalar chirality in similar  models.

\subsection{Entanglement spectra and absence of topological order}
The Li-Haldane conjecture posits that the low-energy sector of the entanglement spectrum of a quantum state is in direct correspondence with the spectrum of the Hamiltonian of states localized to the edge \cite{li2008entanglement}. In a tensor-network framework, this can be made more rigorous: making an entanglement cut in a tensor network representation of a two-dimensional state gives rise to an isometry from the physical degrees of freedom on one side of a cut to the virtual degrees of freedom on the cut's bonds.  Applying this isometry to the reduced density matrix on one side of the cut yields a trace-one operator on the virtual degrees of freedom that we can interpret as a thermal Gibbs state of a theory living on the edge \cite{ciracMatrixProductStates2021}. In infinite cylinder matrix-product states (iMPS) in particular, which form the underlying ansatz of the VUMPS algorithm, this entanglement spectrum can be calculated in sectors of definite momentum around the cylinder using the mixed transfer matrix method of Ref.~\cite{cincioCharacterizingTopologicalOrder2013, tenpy2024}, which gives access to the dispersion relation of the low-energy modes.

The entanglement spectrum has proven a very useful tool to diagnose the presence of topological order---in phases with protected edge modes, either due to chiral central charge or symmetry enrichment, the low-energy part of the entanglement spectrum is the same as the spectrum of the edge theory. 
This edge spectrum is robust and easy to identify numerically because it is separated from entanglement spectrum due the bulk by a gap, related to the bulk excitation gap.  
Since the presence of symmetry-breaking does not necessarily imply the absence of topological order, it is plausible that the chiral FM state has additional topological order, the most likely candidate being a chiral spin liquid (CSL).
To investigate this further, we first consider the entanglement spectrum in VUMPS, which contains distinct signatures if the ground state is in a CSL phase~\cite{cincioCharacterizingTopologicalOrder2013,szasz2020chiral}. 

\begin{figure}
    \centering
    \includegraphics[scale=1.0]{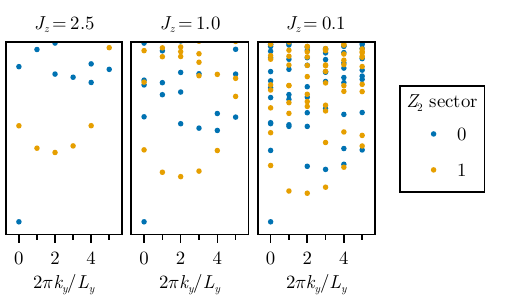}
    \caption{In the FM phase at $J_+ = 1$, the entanglement spectrum demonstrates time-reversal breaking, as it is not symmetric under $k \mapsto -k$.  The entanglement gap gets smaller as the phase gets closer to the tricritical point at $(J_+ = 1, J_z=0)$, but it does not actually become gapless. Here, $L_y=6$ is the cylinder circumference and $k_y$ is the momentum around the cylinder.}
    \label{fig:entspec}
\end{figure}

Entanglement spectra in the FM phase for different values of $J_z$ along $J_+ = 1$ are shown in Fig.~\ref{fig:entspec}. 
Since the FM phase breaks both time reversal and spatial mirror, so we see an expected asymmetry under $k \mapsto -k$. 
The edge spectrum of a state with topological order would have gapless linearly dispersing chiral edge modes; further, for a CSL one should observe a specific count of the number of edge states as a function of momentum around the cylinder~ \cite{szasz2020chiral}.
In our data, the low-energy states in the entanglement spectrum show a gap. 
Additionally, their dispersion appears to be quadratic, indicating that the chiral FM phase does not have the topological order of a CSL. 

To conclusively check for the presence of any topological order in the chiral FM phase, we wrap our cylinder wavefunctions onto a finite torus and calculate their behavior under modular transformations \cite{cincioCharacterizingTopologicalOrder2013}.  
Our results are only consistent with a total quantum dimension of one, ruling out any topological order.

\section{Landau theory for the chiral stripe phase}\label{sec:landau}
In this section, we provide a detailed derivation of the Landau theory for the phase transition from the chiral orbital ferromagnet to the collinear stripe, arguing that an intermediate chiral stripe must intervene if the phase transitions are continuous.
For completeness, let us recall the basic features of our ED calculations. 
The ED results show a ground state that (i) carries non-zero scalar pseudospin chirality (both chiral and anti-chiral states are present and can be resolved by a tiny out-of-plane magnetic field), (ii) is at the $\Gamma$-point, i.e., preserves translation symmetry, (iii) has in-plane antiferromagnetic correlations with all nearest neighbors with peaks at the three $M$ points, (iv) has out-of-plane ferromagnetic correlations, and (v) preserves the internal $\mathbb{Z}_2$ symmetry. 
The goal of this section is to show that there exists a classical ground state predicted by Landau theory (in the thermodynamic limit) which, when appropriately symmetrized over all lattice translations, reproduces the properties of the quantum state. 
Specifically, we will consider classical states with a four-site unit cell, predominantly antiferromagnetic nearest-neighbor correlations in-plane and ferromagnetic correlations out-of-plane. 

To construct the Landau free energy functional $F$, we consider only the soft fields near the $M$ points and the $\Gamma$ point.
Recall that coordinates of the three $M$ points are set by $\M_1 = \b_2/2$, $\M_2 = - \b_1/2$ and $\M_3 = (\b_1 - \b_2)/2$, such that $\M_1 + \M_2 + \M_3 = 0$~[Fig.~\ref{fig:lat}(b)]. 
We note that under the point group generators, the $M$ points transform as $C_{3z} \M_{1,2,3} = \M_{2,3,1}$ and  $M_y \M_{1,2,3} = \M_{1,3,2}$. 
With these conventions, we work out the transformation of the soft-fields, defined in Eq.~\eqref{eq:softfields} in the main text, using the symmetry actions defined in Table~\ref{tab:sym}.
For completeness, we provide the definitions of these soft fields below.
\begin{equation}
\langle \sigma^+_{\r_i} \rangle = \sum_{n=1}^{3} \psi_n \, e^{i \M_n \cdot \r_i}, \; \langle \sigma^z_{\r_i} \rangle = m_0.
\end{equation}
Recall that $\psi_n$ is a complex field which may be thought of as the local orientation of the in-plane component of the spin vector at momentum $\M_n$, while $m_0$ is a real field. 
Using Table~\ref{tab:sym}, we have
\begin{equation}
\langle \sigma^+_{\r_i} \rangle \xrightarrow[]{C_{3z}}  \omega^2 \langle \sigma^+_{C_{3z}(\r_i)} \rangle = \omega^2 \sum_{n=1}^{3} \psi_n e^{i \M_n \cdot C_{3z}(\r_i)} = \omega^2 \sum_{n=1}^{3} \psi_n e^{i C_{3z}^{-1}(\M_n) \cdot \r_i} = \omega^2 \sum_{n=1}^{3} \psi_{n+1} e^{i \M_n \cdot \r_i} \equiv \sum_{n=1}^{3} \psi_n^\prime  e^{i \M_n \cdot \r_i}
\end{equation}
where we have used that $C_{3z}^{-1}(\M_n) = \M_{n-1}$ with $\M_0 \equiv \M_3$. 
This implies that under $C_{3z}$, $\psi_n \to \psi^\prime_n = \omega^2 \psi_{n+1}$, or in matrix notation:
\begin{equation}
\begin{pmatrix}
    \psi_1 \\ \psi_2 \\ \psi_3
\end{pmatrix} \xrightarrow[]{C_{3z}} \omega^2 \begin{pmatrix}
    0 & 1 & 0 \\
    0 & 0 & 1 \\
    1 & 0 & 0 \\
\end{pmatrix} \begin{pmatrix}
    \psi_1 \\ \psi_2 \\ \psi_3
\end{pmatrix} 
\end{equation}
Similarly, under $M_y$, we have
\begin{equation}
\langle \sigma^+_{\r_i} \rangle \xrightarrow[]{M_y} \langle \sigma^{-}_{M_y(\r_i)} \rangle = \sum_{n=1}^{3} \bar{\psi}_n e^{i \M_n \cdot M_y(\r_i)} = \bar{\psi}_1 e^{i \M_1 \cdot \r_i} + \bar{\psi}_3 e^{i \M_2 \cdot \r_i} + \bar{\psi}_2 e^{i \M_3 \cdot \r_i} \equiv \sum_{n=1}^{3} \psi_n^\prime  e^{i \M_n \cdot \r_i}
\end{equation}
where we have used that $\M_1$ is invariant under $M_y$, while $\M_{2}$ and $\M_3$ are swapped by $M_y$.
This implies that under $M_y$, $(\psi_1, \psi_2, \psi_3) \to (\bar{\psi}_1, \bar{\psi}_3, \bar{\psi}_2)$, where $\bar{\psi}_n$ denotes the complex conjugate of $\psi_n$. 
Finally, under time-reversal and $\mathbb{Z}_2$, which are both on-site symmetries, we simply have $\psi_n \to \bar{\psi}_n$ and $\psi_n \to - \psi_n$ respectively. 

For the magnetization density $m_0$, the transformations are simpler.
$C_{3z}$ and $\mathbb{Z}_2$ map $m_0 \to m_0$, while $M_y$ and $\mathcal{T}$ map $m_0 \to - m_0$. 

The symmetry-constrained Landau free energy functional is given by 
\begin{align}
F &= F^{(2)} + F^{(4)}, \nonumber \\
F^{(2)} & = r_0 m_0^2 + r \left( \sum_{n=1}^{3} |\psi_n|^2 \right) + \lambda \, \text{Re}\big[\psi_1^2 + \omega \psi_2^2 + \omega^2 \psi_3^2\big] \nonumber \\
F^{(4)} & = u_0 m_0^4 + u_1 \left( \sum_n |\psi_n|^4 \right) + u_2 \left(\sum_{n < n'} |\psi_n|^2 |\psi_{n'}^2| \right) + u_3 \sum_{n < n'} \left( \text{Re}\big[\bar{\psi}_n \psi_{n'}\big] \right)^2  + w \, m_0^2 \left( \sum_{n=1}^{3} |\psi_n|^2 \right) \nonumber \\
& ~~~ + w_1 \, m_0^2 \, \text{Re}\big[\psi_1^2 + \omega \psi_2^2 + \omega^2 \psi_3^2\big] + F^{(4)}_{\text{non}-O(2)}
\label{eq:GLFapp}
\end{align}
where the first line for $F^{(4)}$ only contains quartic terms consistent with a global internal $O(2)$ symmetry that maps $\psi_n \to \psi_n e^{i \alpha}$~\cite{Jin2025}, whereas the second line contains terms that are not symmetric under these internal $O(2)$ rotations but are allowed under the physical $C_{3v}$ symmetry group. 
Of this latter group of terms, we have explicitly written down the symmetry-allowed quartic coupling between $m_0$ and $\psi_n$ in Eq.~\eqref{eq:GLFapp}.
The additional terms, which involve only $\psi_n$ and break $O(2)$ to the physical $C_{3v}$ symmetry are present in $F^{(4)}_{\text{non}-O(2)}$: for completeness, we note these terms below.
\begin{align}
F^{(4)}_{\text{non}-O(2)} = & \,   v_1 \, \text{Re}[\psi_1^4 + \omega^2 \psi_2^4 + \omega \psi_3^4] + v_2 \, \text{Re}[\psi_2^2 \psi_3^2 + \omega \psi_3^2 \psi_1^2 +  \omega^2 \psi_1^2 \psi_2^2] + v_3 \, \text{Re}[\psi_1^2 |\psi_1|^2 + \omega \psi_2^2 |\psi_2|^2 + \omega^2 \psi_3^2 |\psi_3|^2] \nonumber \\ 
& + v_4 \, \text{Re}[\psi_1^2 (|\psi_2|^2 + |\psi_3|^2) + \omega \psi_2^2 (|\psi_3|^2 + |\psi_1|^2) + \omega^2 \psi_3^2 (|\psi_1|^2 + |\psi_2|^2)] \, .
\end{align}
Each of these terms can be generated by starting with the first quartic term and symmetrizing it with $C_{3z}$, $M_y$ and $\mathcal{T}$.
For the rest of the analysis, we will neglect these additional anisotropic quartic terms in $F^{(4)}_{\text{non}-O(2)}$ for simplicity, i.e., we will set $v_1 = v_2 = v_3 = v_4 = 0$.

Having laid out the most general symmetry-allowed Landau free energy, we now turn to an analysis of its minima. 
First, we will analyze the theory without the ferromagnetic field at the $\Gamma$ point, i.e., we will set $m_0 = 0$ and derive the configurations of $\psi$ which minimize the free energy -- these determine the in-plane components of $\langle \bm{\sigma} \rangle$.
Later, we will argue that having $m_0 \neq 0$ will lead to a non-coplanar spin texture with non-zero chirality, but does not change our main conclusions about the texture of the in-plane components of $\langle \bm{\sigma} \rangle$. 

It is convenient to move to a representation of the order parameter $\psi_n$ in terms of amplitude and phase, i.e,. $\psi_n = \rho_n e^{i \phi_n}$.
This leads to the following expression for the free energy
\begin{align}
F[\{ \psi_n = \rho_n e^{i \phi_n} \}] = & r \sum_n |\psi_n|^2 +  \lambda \, \left[\rho_1^2  \cos\left( 2 \phi_1 \right) + \rho_2^2 \cos\left( 2 \phi_2 - \frac{4\pi}{3}\right) + \rho_3^2  \cos\left( 2 \phi_3 + \frac{4\pi}{3} \right)\right] + u_1 \sum_n \rho_n^4 + u_2 \sum_{n < n^\prime} \rho_n^2 \rho_{n^\prime}^2  \nonumber \\ 
& + u_3 \sum_{n < n^\prime} \rho_n^2 \rho_{n^\prime}^2 \cos^2 \left( \phi_n - \phi_{n^\prime} \right)
\label{eq:FampphaseApp}
\end{align}
We note that at the quadratic level, the terms controlling the phases of $\psi_n$ simply prefer to lock to $\phi_n = 2 (n -1)\pi/3$ if $\lambda < 0$, and to angles shifted by $\pi/2$ if $\lambda > 0$. 
This can be interpreted as the spin direction locking to the lattice.
At the quartic level, however, provided $u_3$ is positive, one prefers a state where the in-plane spin-vectors on the 4 sites of the unit cell are orthogonal, i.e., $|\phi_n - \phi_n^\prime| = \pi/2$. 
For a continuous phase transition, we expect $\rho_n$ to be small near the critical point, so the first scenario is more appropriate.
However, if the phase transition is a fluctuation-driven first-order transition, then $\rho_n$ can immediately pick up a (relatively) large non-zero value upon ordering, and the $u_3$ term can dominate. 
We will argue that a continuous transition prefers a $1M$ (collinear stripe) or a $3M$ (chiral stripe/dense skyrmion crystal) state, although a $2M$ state can potentially be preferred if the phase transition is strongly first order.

In case of a continuous transition, the phases of $\psi_n$ are locked by the quadratic term as discussed above, and the free energy takes a simpler form in terms of a vector $\bm{\rho} = (\rho_1, \rho_2,\rho_3)$:
\begin{equation}
F = (r - |\lambda|) \bm{\rho}^2 + u_1 (\bm{\rho}^2)^2 + \left(u_2 + \frac{u_3}{4} - 2 u_1\right) \sum_{n < n^\prime} \rho_n^2 \rho_{n^\prime}^2
\label{eq:FEamp}
\end{equation}
When $r - |\lambda| < 0$, $\bm{\rho}$ prefers to acquire a non-zero expectation value, and by finding the minima with respect to each component $\rho_n$ (or rather  $\rho_n^2$), we find the following three equations:
\begin{align}
& - |r - |\lambda|| + 2 u_1 \rho_1^2 +  \left(u_2 + \frac{u_3}{4}\right) (\rho_2^2 + \rho_3^2) = 0, \, \text{ and cyclic permutations of this equation.} \nonumber \\ 
& \implies - 3|r - |\lambda|| + \left[ 2 u_1  + 2 \left(u_2 + \frac{u_3}{4}\right) \right] \bm{\rho}^2 = 0 \, \implies \bm{\rho}^2 = \frac{3|r - |\lambda||}{2\left[ u_1 + u_2 + \frac{u_3}{4} \right]}
\label{eq:FEminZeroM}
\end{align}
where in the second step we have added the three equations. 
This clearly indicates that the magnitude of $\bm{\rho}$ is set to a value $\rho_0$ by the Landau parameters, but the relative components are not determined. 
To determine this, we have to look at the full free energy. 
We consider the $1M$ state with $\bm{\rho} = \rho_0(1,0,0)$, the $2M$ state with $\bm{\rho} = \rho_0(1,1,0)/\sqrt{2}$, and the $3M$ state with $\bm{\rho} = \rho_0(1,1,1)/\sqrt{3}$, and compare their free energies (of course, other states are possible but unlikely for reasons we will see below). 
The only term that distinguishes between these three cases is the last term in Eq.~\eqref{eq:FEamp}
\begin{align}
F_{1M} &= (r - |\lambda|) \rho_0^2 + u_1 \rho_0^4  \nonumber \\
F_{2M} &= (r - |\lambda|) \rho_0^2 + u_1 \rho_0^4 + \left(u_2 + \frac{u_3}{4} - 2 u_1\right) \frac{\rho_0^4}{2} \nonumber \\
F_{3M} &= (r - |\lambda|) \rho_0^2 + u_1 \rho_0^4 + \left(u_2 + \frac{u_3}{4} - 2 u_1\right) \rho_0^4
\end{align}
Hence, if $u_2 + u_3/4 > 2 u_1$, i.e., the mutual repulsion between the different fields is stronger, then we prefer the $1M$ state.
By contrast, if $u_2 + u_3/4 < 2 u_1$, i.e., the self repulsion of a field is stronger, then we prefer the $3M$ state. 
Within this analysis, the $2M$ state is never preferred. 

We next consider the classical (pseudo)spin texture of these states. 
The $1M$ state is simply the collinear stripe with a two-site unit cell.
Both the $2M$ and the $3M$ states have four sites within each unit cell.
While the $2M$ state can be classical in the sense that we can have a classical in-plane spin configuration $\psi_1 =  1/\sqrt{2}, \, \psi_2 =  i/\sqrt{2}, \, \psi_3 = 0$ which respects the unit length constraint at every site. 
This state corresponds to a four site unit cell with orthogonal alignment of nearest neighbor spins. 
By contrast, the $3M$ state has a vanishing size of in-plane moment at one of the four sites within the unit cell, indicating that the spin must completely cant out of the plane if the fixed size constraint were to be respected. 
Therefore, the soft-spin nature, potentially arising from quantum fluctuations, is necessary if the 3M state is to be energetically favorable. 

We now extend our discussion to include the soft field $m_0$ corresponding to uniform $S^z$ ordering, that we have neglected thus far. 
If the coupling $w$ is negative (positive), there is attractive (repulsive) interaction between the fields. 
Since the ED observation indicates some possible coexistence, it is reasonable to assume that $w < 0$. 
The minimization of the free energy gives the following equations:
\begin{align}
\frac{\partial F}{\partial (|\psi_n|^2)} &= r + \lambda \cos\left( 2\phi_n - \frac{2(n-1)\pi}{3}  \right) + 2 u_1 |\psi_n|^2 +  \sum_{n^\prime \neq n} \left[ u_2 |\psi_{n^\prime}|^2 +  u_3 |\psi_{n^\prime}|^2 \cos^2\left( \phi_n - \phi_{n^\prime} \right) \right] + w m_0^2 = 0 \nonumber \\
\frac{\partial F}{\partial (m_0^2)} &= r_0 + 2 u_0 m_0^2 + w \sum_n |\psi_n|^2 = 0 
\end{align}
We first note that the equilibrium value of $m_0^2 = - r_0 - w \sum_n |\psi_n|^2$, which indicates a region of coexistence of non-zero $m_0$ and non-zero $\psi_n$ around the decoupled critical point $r_0 = 0$, provided $w < 0$. 
Further, if we substitute the value of $m_0^2 = - r_0 - w \sum_n |\psi_n|^2$ from second equation into the first one, we find a set of equations for $\psi_n$ that resembles the set of equations for $\psi_n$ specified in Eq.~\eqref{eq:FEminZeroM}, but with renormalized coefficients for $r$, $u_1$ and $u_2$. 
\begin{equation}
 \left( r - \frac{w r_0}{2u_0} \right) + \lambda \cos\left( 2\phi_n - \frac{2(n-1)\pi}{3}  \right)  + 2\left( u_1 - \frac{w^2}{4u_0}\right) |\psi_n|^2 +  \sum_{n^\prime \neq n} \left[ \left(u_2 - \frac{w^2}{4u_0} \right) |\psi_{n^\prime}|^2 +  u_3 |\psi_{n^\prime}|^2 \cos^2\left( \phi_n - \phi_{n^\prime} \right) \right] = 0
\end{equation}
Therefore, within the Landau framework, the $1M$ ($3M$) state remains favorable if $u_2 + u_3/4 > 2 u_1$ ($u_2 + u_3/4 < 2 u_1)$, which are identical conditions to what we found previously.  
We note that if $w$ is sufficiently large, then the quartic part of the free energy becomes unbounded from below, and higher order terms will be required to stabilize the system. 
Finally, we also note that the quartic term $w_1 m_0^2 \, \text{Re}[\psi_1^2 + \omega \psi_2^2 + \omega^2 \psi_3^2] $, which also couples the FM and stripe order parameters, will renormalize $\lambda \mapsto \lambda - w_1/2u_0$ in Eq.~\eqref{eq:FampphaseApp}, which does not change any of our conclusions except at the fine-tuned limit $\lambda = w_1/2u_0$.

\vfill

\end{document}